\documentclass[sigconf,authorversion,nonacm,10pt]{acmart}
 \setcopyright{none}

\usepackage{tikz}
\usepackage{amsmath}
\usepackage[noreview]{opera}
\usepackage{algorithmic}
\usepackage[linesnumbered,lined,vlined,ruled,commentsnumbered,noend]{algorithm2e}
\usepackage{enumitem}
\usepackage{url}

\usepackage{tabularx}
\usepackage{caption}
\usepackage{hyperref}
\usepackage{listings}
\usepackage{pgfplots}
\usepackage{balance}

\newcommand*\circled[1]{\raisebox{.5pt}{\textcircled{\raisebox{-.9pt} {#1}}}}

\newcommand{\system}{\textsc{saBPF}\xspace}
\newcommand{\capturesystem}{\textsc{ProvBPF}\xspace}
\newcommand{\ebpf}{eBPF\xspace}
\newcommand{\lsm}{LSM\xspace}
\newcommand{\camflow}{CamFlow\xspace}

\begin{document}

\title{\reviewchange{Secure Namespaced Kernel Audit for Containers}}
\titlenote{This paper was accepted at the 12th ACM Symposium on Cloud Computing (SoCC'21) (see \url{https://acmsocc.org/2021/}).}

\author{Soo Yee Lim}
\affiliation{%
	\institution{University of British Columbia}}

\author{Bogdan Stelea}
\affiliation{%
	\institution{University of Bristol}}

\author{Xueyuan Han}
\affiliation{%
	\institution{Harvard University}}

\author{Thomas Pasquier}
\affiliation{%
	\institution{University of British Columbia}}

\renewcommand{\shortauthors}{Lim \etal}

\date{}

\begin{abstract}
	Despite the wide usage of container-based cloud computing, 
  container auditing for security analysis relies mostly on
  built-in host audit systems, which
  often lack the ability to capture high-fidelity container logs.
State-of-the-art reference-monitor-based audit techniques
  greatly improve the quality of audit logs, but their system-wide
  architecture is too costly to be adapted for individual containers.
Moreover, these techniques typically require extensive kernel
  modifications, making it difficult to deploy in practical settings.

In this paper, 
  we present \system \reviewchange { (\textbf{s}ecure \textbf{a}udit \textbf{BPF})}, an extension of the \ebpf framework 
  capable of deploying
  secure system-level audit mechanisms at the container granularity.
We demonstrate the practicality of \system in Kubernetes 
  by designing an audit framework, an intrusion detection system,
  and a lightweight access control mechanism.
We evaluate \system and show that
  it is comparable in performance and security guarantees
  to audit systems from the literature
  that are implemented directly in the kernel.

\end{abstract}

\maketitle

\section{Introduction}
\label{sec:introduction}
In recent years, 
 container-based cloud computing has gained much traction.
As a lightweight alternative to VM-based computing
  infrastructure, it provides an attractive multi-tenant environment
  that supports the development of microservice architecture,
  where a monolithic application is organized into a number of
  loosely-decoupled services for modularity, scalability, and fault-tolerance~\cite{soltesz2007container}.

Container security becomes a major concern as the popularity of
  container-based cloud continues to grow rapidly~\cite{containerstate}. 
For example, by sharing individual microservices across applications,
  the container ecosystem, as promoted by widely-used
  container management and orchestration platforms, such as Docker
  and Kubernetes, inadvertently spreads vulnerabilities that can widen
  an application's attack surface~\cite{torkura2017integrating}.
Vulnerabilities in individual containers can facilitate the construction of
  a \emph{cyber kill-chain}, in which the attackers perform various attacks
  in steps on different microservices to achieve their ultimate goal~\cite{jin2019dseom}.
Information leakage between a host and a container and between two
  co-resident containers has also been demonstrated to be possible~\cite{gao2017containerleaks}.
  
Like in traditional security analysis, system-layer audit logs are often
  considered to be an important information source for addressing many 
  container security concerns~\cite{han2020ndss, han2020sigl, milajerdi2019poirot}.
For example, to identify a misbehaving container from a cluster of
  replicated microservices, kernel audit logs have been used
  to define process activity patterns and describe unusual activity
  that does not fit into any observed pattern~\cite{hassan2018towards}.
Container-focused security systems typically use existing
  host audit tools, such as the Linux Audit Framework, to log system
  events, but research has shown that these audit systems are insufficient to
  capture complete system activity necessary for security analysis~\cite{gehani2012spade}.
Alternative \emph{reference-monitor-based} approaches~\cite{pasquier2017practical-socc2017,  pohly2012hi} provide a more complete picture by leveraging in-kernel monitoring hooks, 
but they are not designed with container-based computing architecture in mind.
Specifically, reference-monitor-based approaches require extensive host kernel modification and permit only
  host-wide policy specification.
The former requirement is often forbidden by cloud infrastructure providers,
  and the latter makes it difficult to satisfy the audit needs of individual
  containers sharing the same host.

We present \system, a lightweight, secure kernel audit system
  for containers. 
\system enables each container to customize its auditing mechanism and policy, even if
  containers specifying different policies and mechanisms are co-located on the same host.
Audit data captured by \system is guaranteed to have \emph{high fidelity},
  meaning that the data \emph{faithfully} records \emph{complete} container-triggered system activity~\cite{pohly2012hi}, 
  free of concurrency vulnerabilities~\cite{watson2007exploiting} and 
  missing records~\cite{gehani2012spade} 
  that are commonly present in existing audit tools.
As such, \system builds a solid foundation for future forensic applications in
  a containerized cloud environment.
For example, to deploy a cyber kill-chain detection system for a network of
  Kubernetes containers (or \emph{pods}), we can follow the \emph{Sidecar}
  design pattern, in which \system is configured for each pod based on the
  characteristics of the microservice it provides.
A specialized sidecar
  container is attached to capture and analyze audit logs.
Sidecar containers from different pods can ship any suspicious events
  to a remote system where alert correlation is performed to detect the presence
  of a kill-chain~\cite{milajerdi2019holmes}.
We discuss this use case in more detail in~\autoref{sec:uc}.

\system is implemented as an extension of \ebpf.
Specifically, we make the following contributions:
\begin{itemize}[noitemsep]
	\item We expand Linux's \ebpf framework to support 
	the attachment of \ebpf programs 
	at the intersection of the reference monitor and namespaces, 
	which allows fully-configurable, high-fidelity, system-level auditing
	for individual containers \reviewchange{(see \autoref{sec:extending})};
	\item We develop functional proof-of-concept applications using
	\system to demonstrate its practicality \reviewchange{(see \autoref{sec:uc})};
	\item We conduct thorough performance analysis to understand the cost and benefits of using \system for secure audit; we show that
	our approach outperforms existing audit solutions of similar caliber \reviewchange{(see \autoref{sec:lsm-bpf} and \autoref{sec:evaluation})};
	\item We open source \system to facilitate the development of security applications for containers in the cloud \reviewchange{(see \autoref{sec:availability})}.
\end{itemize}

\section{Background}
\label{sec:background}
\begin{figure}[t]
	\includegraphics[width=\columnwidth]{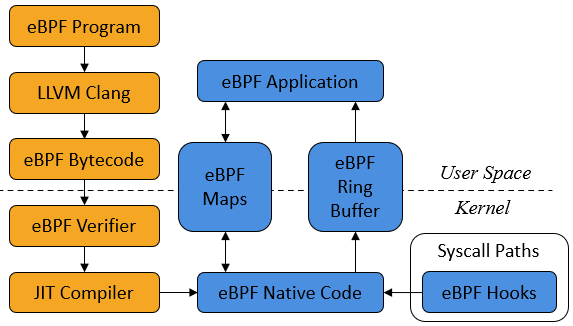}
	\caption{\ebpf workflow.}
	\label{image:ebpf}
\end{figure}

\system is built upon the \emph{extended Berkeley Packet Filter} (\ebpf) framework, 
  a Linux subsystem that allows user-defined programs 
  to safely run inside the kernel~\cite{ebpf_io}.
To provide high-fidelity system audit data,  \system implements \ebpf programs
  that instrument in-kernel hooks defined by the \emph{Linux Security Modules}
  (LSM) framework, which is the reference monitor implementation
  for Linux~\cite{morris2002linux}.
\system further leverages \emph{namespaces} to ensure that auditing can be customized
  for individual containers.
We provide some background knowledge for each component.

\subsection{Extended Berkeley Packet Filter}
\label{sec:background:ebpf}

\ebpf is a Linux built-in framework that allows customized 
  extensions to the kernel
  without modifying the kernel's core trusted codebase.
We illustrate how \ebpf works in \autoref{image:ebpf}.
Developers can write an \ebpf program in C and compile
  the program into \ebpf bytecode using Clang.
The kernel uses a verifier to statically analyze the bytecode,
  minimizing security and stability risks
  of running untrusted kernel
  extensions~\cite{gershuni2019simple}.
After verification, a just-in-time (JIT) compiler
  dynamically translates the bytecode into
  efficient native machine code.
The translated \ebpf program is
  attached to designated kernel locations
  (\eg LSM hooks in our case)
  and executed at runtime.
\ebpf programs can share data with user-space
  applications using special data structures such as \ebpf maps
  and ring buffers~\cite{bpf-ringbuffer}.

\ebpf is frequently used as the underlying framework for network security and
  performance monitoring. 
In a container environment, for example,
  \ebpf enables Cilium~\cite{cilium}, a popular network monitoring tool for 
  platforms such as Kubernetes, to secure application-level protocols
  with fine-grained firewall policies.

\subsection{Linux Security Modules}

\begin{figure}[t]
	\includegraphics[width=\columnwidth]{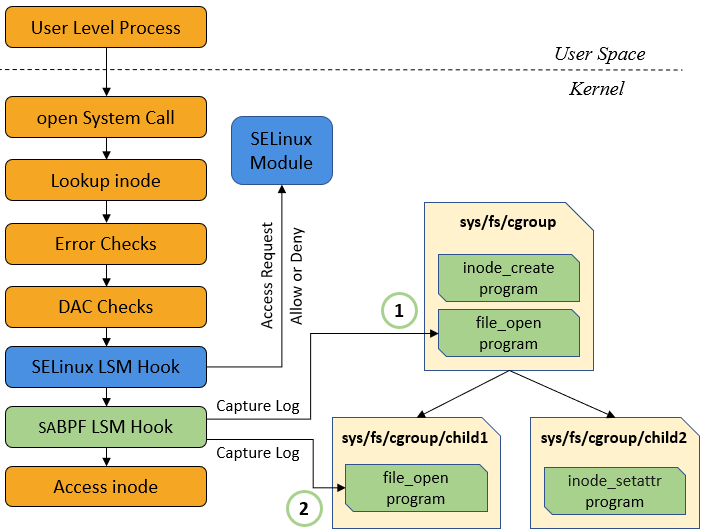}
	\caption{LSM hook architecture~\cite{morris2002linux}. 
	The green blocks are specific to \system, which is described in~\autoref{sec:extending}.
	\reviewchange{\circled{1} shows programs attached to the root cgroup; \circled{2} shows two programs attached to the \texttt{child1} and \texttt{child2} cgroup respectively.}}
	\label{image:lsm}
\end{figure}

LSM consists of a set of in-kernel \emph{hooks} that are strategically placed
  where kernel objects (such as processes, files, and sockets)
  are being accessed, created, or destroyed.
LSM hooks can be instrumented to enable diverse security functionality,
  with the canonical usage being the implementation of mandatory access control scheme~\cite{schreuders2011empowering}.
As a reference monitor, LSM has also been adapted to perform secure kernel logging, 
  which provides stronger \emph{completeness} and \emph{faithfulness} guarantees
  than traditional audit systems~\cite{pasquier2017practical-socc2017, pohly2012hi, bates2015trustworthy}.
For example,  prior research has verified that LSM hooks capture all meaningful interactions between kernel objects~\cite{edwards2002runtime,jaeger2004consistency} and that information flow within the kernel 
  can be observed by at least one LSM hook~\cite{georget2017verifying},
  which is necessary to achieve completeness.

The LSM framework does not use system call interposition as older systems did.
Syscall interposition is susceptible to concurrency vulnerabilities,
  which in turn lead to time-of-check-to-time-of-use (TOCTTOU) attacks
  that result in discrepancies between the events as seen by the security mechanism and the system call logic~\cite{watson2007exploiting,watson2013decade}.
This is why solutions such as kprobe-BPF, while useful for performance analysis, are not appropriate to build security tools.
Instead, LSM's reference-monitor design ensures that the relevant kernel states and objects are immutable
  when a hook is triggered, which is necessary to achieve faithfulness.
\reviewchange{LSM-BPF~\cite{krsi_lwn} is a recent extension to the \ebpf framework
  that provides 
  a more secure mechanism to implement 
  security functionalities on LSM hooks. 
}

\subsection{Namespaces in Linux}
\label{sec:background:namespaces}

\begin{table}[]
	\begin{tabularx}{\columnwidth}{l|X} 
		\textbf{Name} & \textbf{Description} \\ \hline
		cgroup        & Allocate system resource (\eg CPU, memory,  and networking) \\\hline
		ipc               & Isolate inter-process communications \\\hline
		network      & Virtualize the network stack \\\hline
		mount         & Control mount points \\\hline
		process       & Provide independent process IDs \\\hline
		user             & Provide independent user IDs and group IDs, and 
		                       give privileges (or capabilities) associated with 
		                       those IDs within other namespaces \\\hline
		UTS             & Change host and domain names \\\hline
		Time            & See different system times                
	\end{tabularx}
	\vspace{1mm}
	\caption{Summary of Linux namespaces.}
	\label{table:namespaces}
\end{table}

A namespace in the Linux kernel is an abstract environment in which
  processes within the namespace appear to own an independent instance
  of system resources.
Changes to those resources are not visible to processes outside the namespace.
We summarize available namespaces in Linux in \autoref{table:namespaces}.

\begin{figure}[t]
	\includegraphics[width=0.7\columnwidth]{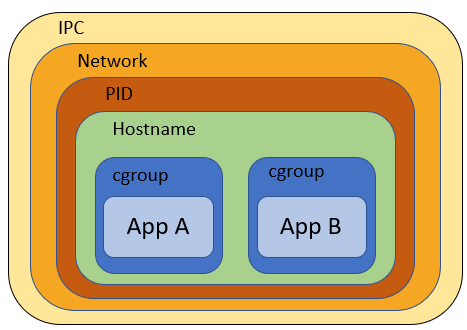}
	\caption{Namespaces in Kubernetes.}
	\label{fig:namespaces}
\end{figure}

One prominent use of namespaces is to create containers.
For example, an application in a Docker container runs within its own set of namespaces.
Kubernetes ``pods'' contain one or more containers so that they share
  namespaces (and therefore system resources).
Kubernetes makes it appear to applications within a pod that 
  they own a machine of their own (\autoref{fig:namespaces}).

\system modifies the kernel to enable per-container auditing, 
  selectively invoking \ebpf programs on LSM hooks based on \texttt{cgroup} membership (\autoref{sec:namespacing}).
\texttt{cgroup} isolates processes' resource usage in a hierarchical fashion, 
  with a child group having additional restrictions to those of its parent.
Since \texttt{cgroup} v2~\cite{cgroupv2}, 
  this hierarchy is system-wide, and all processes initially belong to the root \texttt{cgroup}.
In a Kubernetes pod, for example, containers can be organized in a hierarchical structure and assigned various \texttt{cgroup} namespaces to set up further restrictions.
Certain types of \ebpf programs, such as the ones that are socket-related,
  can already be attached to \texttt{cgroup}s (\eg \texttt{BPF\_PROG\_TYPE\_CGROUP\_SKB}). 
This allows,  for example, a packet filtering program to apply network filters
  to sockets of all processes within a particular container.
\system makes it possible to attach \ebpf programs 
  at the intersection of \texttt{cgroup}s and LSM hooks
  (\autoref{sec:namespacing}) for audit purpose and beyond (\autoref{sec:uc}).

\section{\system: Extending the \ebpf Framework}
\label{sec:extending}
\system extends the \ebpf framework to support
  secure kernel auditing in a containerized environment.
Our design is \emph{minimally invasive}, reusing
  existing components in the framework as much as possible and
  extending only what we deemed to be necessary.
This is a conscious decision made to achieve two objectives:
1) by adhering closely to the design philosophy of the \ebpf framework,
  we ensure that \system can be readily integrated into the mainline kernel;
2) since \system is built upon the \ebpf framework and
  adheres to its design philosophy, users already familiar with \ebpf can quickly
  develop new applications using \system, while new users have access to
  {\ebpf}'s documentations and forums, which makes \system easy to learn and use.
In addition,  any existing \ebpf program can run
  in conjunction with our audit solution on the \system-enhanced platform
 for individual containers.

\subsection{Namespacing LSM-BPF}
\label{sec:namespacing}

\system extends the use of \texttt{cgroup}
   (which in \ebpf is used mostly for network filtering)
 to LSM hooks.
This extension allows \system to precisely control audit granularity.
Recall that \texttt{cgroup}s are
  arranged in a global hierarchy since \texttt{cgroup} v2, and all processes belong to the root \texttt{cgroup} by default.
A Kubernetes pod, for example, defines a \emph{pod-level} \texttt{cgroup},
  which is the ancestor of \emph{container-level} \texttt{cgroup}s within which
  individual containers in the pod reside (\autoref{fig:namespaces}).
By attaching \ebpf programs to container-level \texttt{cgroup}s,
  \system can perform container-level auditing;
at the same time,
  \system can monitor activity inside the entire pod by
  attaching audit programs to the pod's root \texttt{cgroup}.
While the design of Kubernetes makes it natural to follow this two-level
  audit scheme using \texttt{cgroup},
\system can support arbitrarily complex \texttt{cgroup} hierarchy for customizable use.

\autoref{listing:cgprogram} illustrates how an \system program is defined by developers.
It is a program that simply prints ``Hello World!'' when the \texttt{file\_open} LSM hook is triggered.
The statement in line 1 specifies where the program should be attached, and the \texttt{ctx}
  variable in line 2 contains the parameters passed from the \texttt{file\_open} hook.
The user attaches (and detaches) this program to (and from) a \texttt{cgroup}
  through the \texttt{bpf()} system call.
Multiple \ebpf programs can be attached to the same \texttt{cgroup}-hook pair
  by setting the \texttt{BPF\_F\_ALLOW\_MULTI} flag; they will be executed in FIFO order.
  
\begin{lstlisting}[float=t, numbers=left, language=C, style=customC, caption={A ``Hello World!'' \system program that can be triggered on the \texttt{file\_open} hook.}, label=listing:cgprogram]
SEC("cgroup_lsm/file_open")
int cgroup_file_open(struct bpf_cgroup_lsm_ctx *ctx)
{
  bpf_trace_printk("Hello World!\n");
	return 0;
}
\end{lstlisting}

In the kernel, when an LSM hook is triggered, \system invokes appropriate \ebpf programs
  through a customized security module.
This module performs three main actions for every hook.
First, it prepares the parameters, which are passed from the hooks to the \ebpf programs.
It then retrieves the \texttt{cgroup} associated with the current task.
Finally, it traverses the \texttt{cgroup} hierarchy (\autoref{sec:background:namespaces}) backwards
  from the current \texttt{cgroup} to the root \texttt{cgroup}
and executes all programs associated with the LSM hook.

\autoref{image:lsm} illustrates this process using the
  \texttt{open} system call as an example.
\reviewchange{
The root \texttt{cgroup} has two child \texttt{cgroup}s, 
\texttt{child1} and \texttt{child2}.
While the root \texttt{cgroup} has programs
attached to both \texttt{inode\_create} and \texttt{file\_open} LSM hooks,
\texttt{child1} has only one program
attached to \texttt{file\_open} and
\texttt{child2} has one program
attached to a different LSM hook, \texttt{inode\_setattr}.
As a result,
\emph{any} process that triggers the
\texttt{file\_open} hook
leads to the invocation of the programs
attached to that hook in the root \texttt{cgroup}.
However, the programs attached to the same hook
in \texttt{child1} are only called
if a process belonging to \texttt{child1} (or one of its decedents)
triggers the hook.
Note that this process still causes
programs in the root \texttt{cgroup} to be called.
}

Early in the design phase, we considered creating a dedicated namespace
  for system auditing. %
While this allows a clear separation of namespaces' responsibilities,
  given that \texttt{cgroup}s are designed to control access to system resources,
we eventually abandoned this design for two reasons.
First, significant re-engineering of existing container solutions would be required to make
  use of this new namespace.
Second, existing namespaced \ebpf already uses \texttt{cgroup}.
  We believe that introducing a new namespace goes against the current design philosophy of \ebpf.
However, we emphasize that based on our experience,
  it is relatively straightforward to implement a new namespace should such a need
  arise in the future.

\subsection{Local Storages}
\label{sec:localstorage}

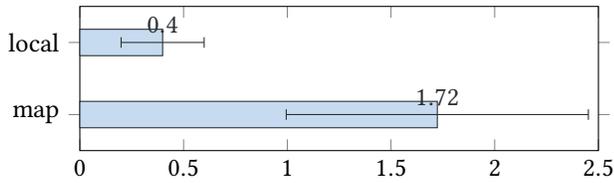
\begin{figure}[t]
	\definecolor{blues1}{RGB}{198, 219, 239}

\begin{tikzpicture}
	\begin{axis}[
		xbar, xmin=0,
		width=\columnwidth, height=3.5cm, enlarge y limits=0.5,
		symbolic y coords={map,local},
		ytick=data,
		xmax=2.5,
		nodes near coords, nodes near coords align={vertical},
		]
		\addplot[
		,blues1!20!black,fill=blues1,
		error bars/.cd,
		x dir=both,x explicit
		] coordinates {
			(0.3990769231,local) +-(0.2001401432,0.2001401432)
			(1.723307692,map) +-(0.7282697971,0.7282697971)
		};
	\end{axis}
\end{tikzpicture}
	\caption{Look-up time for the \texttt{cred} local storage and the \ebpf map in $\mu$s. Using local storage gives a \emph{4x} speedup.}
	\label{image:localstorage:perf}
\end{figure}

System auditing often requires associating data with kernel objects~\cite{pohly2012hi}.
\reviewchange{
In early prototypes, we considered using \ebpf maps, 
which are key-value stores shared 
among multiple \ebpf programs across execution instances, 
but we abandoned the idea due to poor maintainability.
Specifically,
when using \ebpf maps, 
developers must create an entry
for each new kernel object
to store data associated with the object.
The key to the entry must be unique
during the lifecycle of the object.
Ensuring uniqueness
for all kernel objects is important, but prone to error.
For example,
it is insufficient
to use just an inode number
as the key for an \texttt{inode} object;
rather,
a \emph{combination} of the inode number
and the file system's unique identifier is needed
because inode numbers are guaranteed to be unique
\emph{per file system} only.
Moreover,
developers must also take special care
to remove map entries
when objects reach the end of their lifecycle.
This problem is exacerbated by the fact that
\ebpf maps are created with capacity limits.

\system uses a completely different approach
to storing such data,
extending \ebpf's \emph{local storages}, 
which are data structures that are directly
associated with kernel objects.
Local storages provide an interface 
similar to \ebpf maps, 
but they use the object reference
as the key and store the value locally with the kernel object.
At the end of an object's lifecycle 
when the object no longer has any reference,
\ebpf transparently removes the local storage associated with the object.
This takes the responsibility of removing unused entries away from developers, making it less error-prone. 
Furthermore, local storages incur less performance overhead compared to \ebpf maps, as shown in~\autoref{image:localstorage:perf}. 
At the time of writing, \ebpf provides local storages for only \texttt{cgroup},
\texttt{socket}, \texttt{inode} and \texttt{task}.
We implemented additional local storages for \texttt{file}, \texttt{cred}, \texttt{ipc}, \texttt{superblock},
and \texttt{msg\_msg} to \emph{fully} support LSM-based auditing.
We give a practical illustration in \autoref{sec:usecase:provbpf}.
}

\subsection{Extension of \ebpf interface}
\label{sec:helpers}

\begin{table}[]
	\begin{tabularx}{\columnwidth}{p{0.45\columnwidth}|X}
		\textbf{Name} & \textbf{Description} \\ \hline
		\texttt{bpf\_inode\_from\_sock}        & Retrieve~the~inode~associated with a socket \\\hline
		\texttt{bpf\_file\_from\_fown}               & Retrieve~the~file~associated with a \texttt{fown\_struct} \\\hline
		\texttt{bpf\_dentry\_get}      & Retrieve~the~\texttt{dentry}~associated with an inode \\\hline
		\texttt{bpf\_dentry\_put}         & Release a \texttt{dentry} after use \\\hline
		\texttt{bpf\_[cred/msg/ipc/ file]\_storage\_get} & Get a \texttt{bpf\_local\_storage} from a \texttt{cred}/\texttt{msg}/\texttt{ipc}/\texttt{file} \\\hline
		\texttt{bpf\_[cred/msg/ipc/ file]\_storage\_delete} & Delete~a~\texttt{bpf\_local\_storage} from a \texttt{cred}/\texttt{msg}/\texttt{ipc}/\texttt{file}
	\end{tabularx}
  \vspace{1mm}
	\caption{Summary of new \ebpf helpers provided by \system.}
	\label{table:helpers}
\end{table}

To access kernel data, \ebpf programs rely on \emph{\ebpf helpers},
  which are an \reviewchange{allowlist} of kernel functions permitted by the \ebpf verifier
  to interact with the kernel.
\system defines a number of extra \ebpf helpers, as shown in~\autoref{table:helpers},
  to facilitate system auditing.

A subset of functions return the inode associated
  to an object of a certain type (\eg \texttt{bpf\_inode\_from\_sock} returns the inode of
  a socket object).
This can be useful %
  to understand the interplay of system calls
  acting at different levels of kernel abstraction.

\texttt{bpf\_dentry\_get} returns the directory entry of an inode,
  which helps \system programs to retrieve the path associated with the inode.
A directory entry is protected by a reference counter when a program manipulates it.
The reference counter is increased when \texttt{bpf\_dentry\_get} is called and
  \emph{must} be decreased by calling \texttt{bpf\_dentry\_put}
  once the entry is no longer used.
To ensure correctness, we also modified the \ebpf verifier to verify that every \texttt{bpf\_dentry\_get} has a corresponding \texttt{bpf\_dentry\_put} being called on the same code path.

The remaining helpers are used to manipulate local storages (\autoref{sec:localstorage}).
We extended \ebpf map helpers so that
userspace programs can interact with those storages.
In \ebpf maps,
userspace programs can access a set of helpers
via system calls to
\eg update or lookup map entries.
Our extension provides similar support for local storages:
  programs can manipulate data in the local storage of a particular kernel object using appropriate userspace identifiers.
For example, assuming appropriate privileges,
  we can lookup, update, and delete data in a \texttt{cred}\footnote{\texttt{cred} is the credential information associated with a process.}
  object's local storage via its \texttt{PID}.

\section{Use Cases}
\label{sec:uc}
A framework like \system is only useful if it is both \emph{practical} 
  and \emph{performant}.
We discuss three meaningful use cases that we have implemented to
  demonstrate the types of application that \system can easily support,
  showcasing its practicality.
We evaluate \system's performance in~\autoref{sec:evaluation}.

\subsection{Whole-system Provenance Capture}
\label{sec:usecase:provbpf}
We describe our implementation of \capturesystem, 
  a provenance capture mechanism that we developed atop \system.
Provenance has gained much traction in the security community, 
  notably with applications designed to understand intrusions in a computer system~\cite{milajerdi2019poirot,hassan2020tactical,hassan2020omegalog},  prevent data exfiltration~\cite{bates2015trustworthy}, 
  and detect attacks~\cite{manzoor2016fast,han2017frappuccino,han2020ndss,han2020sigl,wang2020you,milajerdi2019holmes}.
\capturesystem captures provenance at the thread granularity, 
  recording information such as security context, namespace, and performance metrics.

\begin{figure}[t]
	\includegraphics[width=\columnwidth]{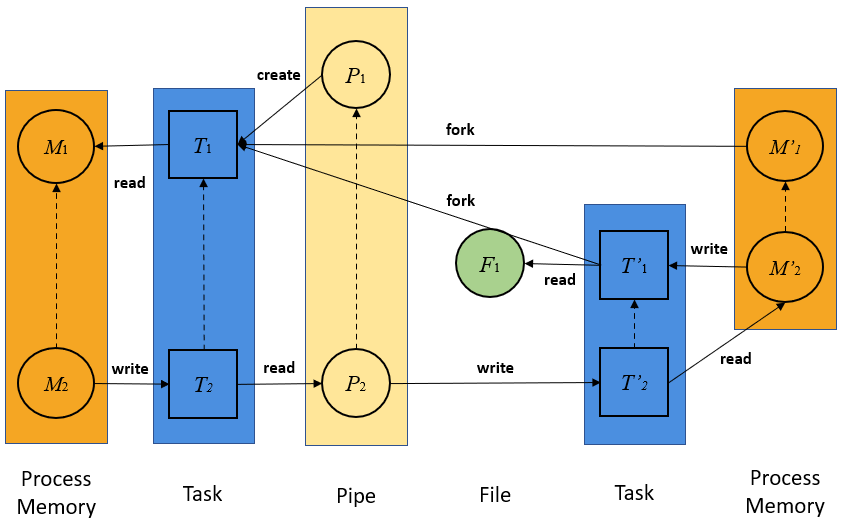}
	\caption{A simplified whole-system provenance subgraph.}
	\label{image:prov-graph-versions}
\end{figure}

\noindgras{A Brief Provenance Introduction.}
Computing systems are too often opaque:
they accept inputs and generate outputs, 
  but the visibility of their inner workings is at best partial,
  which poses many issues in fields ranging from algorithmic transparency 
  to the detection of cybersecurity threats.
Unfortunately, traditional tracing mechanisms are inadequate in addressing these issues.
Instead,
  \emph{whole-system provenance}~\cite{pohly2012hi}, which
  describes system execution by representing information flows 
  within and across systems as a directed acyclic graph, shows promise.
Provenance records subsume information contained in a traditional trace, 
  while causality relationships between events can be inferred through graph analysis.

\autoref{image:prov-graph-versions} shows a simple provenance graph.
\reviewchange{In this graph,
two tasks ($T$ and $T'$) are associated with 
their respective memory ($M$ and $M'$).
The subscripts (\eg $T_1$ and $T_2$) represent 
different \emph{versions} of the same kernel object
to guarantee graph acyclicity~\cite{muniswamy2006provenance}.
$T$ creates a pipe $P$ and 
forks a new process (corresponding to $T'$ and $M'$).
$T'$ reads information from a file $F$ 
and writes information to $P$.
$T$ then reads from this pipe.
A versioned node is created
every time an object receives external information 
(\eg when a task reads from a file).
This is a small subgraph representing a very simple scenario.}
In practice,  for example, a
 graph representing the compilation of the Linux kernel would contain
  approximately a few million graph elements~\cite{moyer2016high}.

The rest of our discussion focuses primarily on
  the novel aspects of {\capturesystem}
  and the design choices we made
  as the result of using \system to capture OS-level provenance
  (instead of modifying the kernel).
We compare {\capturesystem}'s performance to that of a state-of-the-art 
  provenance capture system, \camflow~\cite{pasquier2017practical-socc2017}, in~\autoref{sec:evaluation}.

\begin{figure}[t]
	\includegraphics[width=1\columnwidth]{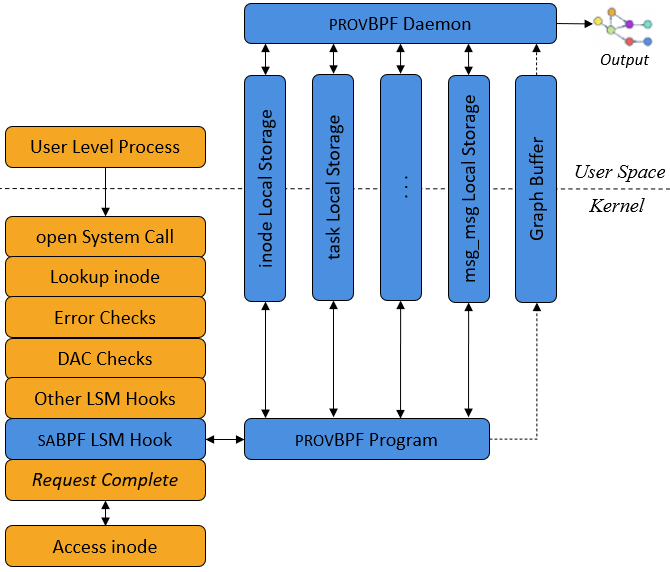}
	\caption{\capturesystem overview.}
	\label{image:overview}
\end{figure}

\noindgras{Overview.}
\autoref{image:overview} illustrates the architecture of \capturesystem
  using the \texttt{open} system call as an example.
In \capturesystem, \ebpf programs are executed on LSM hook invocations.
{\capturesystem}'s \ebpf programs generate provenance graph
  elements in binary and write them to an \ebpf ring buffer~\cite{bpf-ringbuffer}.
A user-space daemon serializes those graph elements and
  outputs them to disk (or to remote endpoints like Apache Kafka~\cite{kafka})
  in a machine readable format such as W3C PROV-DM~\cite{w3cprovdm}.

We must associate \emph{states} with kernel objects
  to guarantee graph acyclicity and to implement graph
  compression algorithms.
\reviewchange{For example, to guarantee acyclicity,
we associate with each object a version counter 
which is updated when external information 
flows into an object and modifies its state.
After the update,
a new vertex is added to the graph
and connected to the previous version of the object 
through a \texttt{version} edge, 
as illustrated in \autoref{image:prov-graph-versions}
as dashed lines.}

We also associate an ``opaque'' flag to the state of certain kernel objects;
opaque objects are not audited.
This is particularly useful for {\capturesystem}'s daemon-related objects,
  because capturing their provenance would result in an
  infinite feedback loop.
\reviewchange{CamFlow uses \texttt{security blobs} from the LSM framework associated with each kernel object to maintain its associated states.
In \capturesystem, we leverages the
  \emph{local storage} mechanism (\autoref{sec:extending}) for this purpose.}
Local storage can be accessed by the \capturesystem daemon from userspace 
  to set a policy for each individual object (\eg to set opacity).

\noindgras{Graph Reduction.}
{\ebpf}-based provenance capture offers 
  exceptional flexibility in designing customized capture policies
  that fulfill different objectives.
Customization typically involves \emph{filtering}, \ie
  selecting kernel objects and system events that are relevant to a specific analysis.
\reviewchange{For example, Bates \etal~\cite{bates2015take} 
only record events related to objects associated with specific SELinux policies.
\capturesystem allows for the filter logic to be built-in during compilation,
  thus reducing run-time overhead.}

\capturesystem implements additional graph reduction techniques
  other than filtering.
It automatically merges consecutive events of the same type between two entities
  into a single event and avoids object versioning as much as possible.
Event merging reduces the number of edges between two nodes
  without changing the semantics of the interactions they represent.
For example, when a process reads a file piece-by-piece through
  a number of successive \texttt{read} system calls, \system would create
  only one directed edge between the process and the file to capture
  these read events, which is sufficient to describe the information
  flow from the file to the process due to read.
On the other hand, avoiding object versioning reduces
  the number of nodes, and \capturesystem does so 
  only when the semantics of an object have not changed.
These graph reduction techniques are completely agnostic to
  specific downstream provenance analysis and can be easily
  configured at compilation time according to the needs of a particular application.
More importantly,
unlike previous work~\cite{Tang:2018:NTB:3243734.3243763}
  that performs graph compression as a costly post-processing step
  (\ie after recording the original graph), 
  \capturesystem employs those techniques
  during capture \emph{before} new edges are added to the graph.

\noindgras{Verifying Capture Correctness.}
\label{sec:evaluation:verifying}
It is challenging to verify the correctness of a provenance capture mechanism~\cite{chan2019provmark}.
At a minimum, we must show that a provenance graph describing system activity
  of a system call makes ``intuitive'' sense for a human analyst inspecting the graph.
We use both static and dynamic analysis to verify that provenance graphs
  generated by \capturesystem are reasonably correct.

Our static analysis generates a graph motif for each system call, 
  which enables us to reason about the semantics of the graph 
  based on our understanding of the system call.
We follow the same strategy as described by Pasquier \etal~\cite{pasquier2018ccs}.
To generate a system call graph motif,
  we first analyze the kernel codebase to construct a call graph of a system call 
  and extract a subgraph, within the call graph, 
  that contains only LSM hooks~\cite{georget2017verifying}.
We then analyze \capturesystem's codebase to generate a graph motif for each 
  LSM hook and augment the subgraph from the previous step by replacing each LSM
  hook in the subgraph with the corresponding graph motif.
The resulting subgraph -- now containing only graph motifs of LSM hooks -- 
  is the graph motif of the system call that summarizes 
  what the provenance graph would look like when the system call is executed.

In addition, we build test programs and follow the same steps above to create
  program-level graph motifs.
\reviewchange{For each program binary,
we build a call graph which we then transform 
into a syscall-only graph.
We replace the syscalls in the graph 
with the motifs we previously built.}
We run each test program and verify by inspection that our (statically-produced) 
  motif matches the (dynamically-produced) provenance graph 
  generated by \capturesystem.
We perform the same steps in \camflow~\cite{pasquier2017practical-socc2017} 
  to verify that the graphs generated by the two systems are equivalent.

\subsection{An Intrusion Detection System for Kubernetes}
\label{sec:usecase:unicorn}
{\system}-based audit systems such as \capturesystem can be used
  as an underlying framework for various security applications in the cloud.
We demonstrate this feasibility through a concrete use case of deploying
  Unicorn~\cite{han2020ndss},  a state-of-the-art host-based
  intrusion detection system (IDS), in a Kubernetes pod using \capturesystem
  as an upstream information provider.
Unicorn is an anomaly-based IDS that learns system behavior from
  the provenance graph generated by benign system activity.
Once a model is learned from the graph, detection is formulated as a graph comparison problem:
  if a running system's provenance graph deviates significantly from the model,
  Unicorn considers the system to be under attack.
In the remainder of this section, we focus our discussion on how \capturesystem
  facilitates deployment of an IDS in a containerized environment in a novel and elegant
  manner; in-depth evaluation of the performance of such an IDS is out of scope and left
  for future work.

\noindgras{Design \& Implementation.}
\capturesystem makes it easy to run a provenance-based IDS
  at the pod level in Kubernetes, which is challenging when provenance data
  is provided by a reference-monitor-based audit system such as CamFlow~\cite{pasquier2017practical-socc2017}.
For systems like CamFlow, provenance is always captured \emph{system-wide};
  as a result, audit logs must be filtered to provide as input to the IDS provenance data relevant
  to a pod only, and filtering must be done on an individual pod basis.
Unfortunately, this extra filtering step inevitably adds delay and complexity to the entire
  detection pipeline, thus reducing \emph{runtime} detection efficacy.

Instead, we use \capturesystem and Kubernetes' sidecar design pattern to
  attach the IDS to Kubernetes applications.
A sidecar container is a container that runs alongside a main container (\ie the one that
  provides core functionality) in a pod.
In our design, for each pod, we include a sidecar container that runs 
  both \capturesystem and Unicorn.
\capturesystem audits the entire pod
  and generates a pod-level provenance graph; the graph is then used as input to
  Unicorn.
We note that other detection systems such as StreamSpot~\cite{manzoor2016fast} and
  log2vec~\cite{liu2019log2vec} could be used in a similar fashion.
In a microservice environment, any misbehavior detected within a single pod can be sent 
  to a dedicated central service that performs alert correlation~\cite{milajerdi2019holmes} 
  to detect, for example, early stages of a cyber kill-chain.

\noindgras{Discussion.}
This deployment strategy, made possible by \capturesystem, have a number of advantages.
First, we do not need to modify applications to deploy our IDS thanks to the sidecar pattern.
Second, we can easily deploy an IDS model specific to an application running in a pod,
  without taking into consideration extraneous activity of the rest of the system.
Third, \capturesystem produces provenance graph elements that can be analyzed directly,
  without introducing filtering delays in the detection pipeline.
Fourth,  deploying \capturesystem imposes no cost on other pods running on the same machine, 
  since \system programs are only triggered within the context of a single pod.
This is in contrast to a classic system-wide approach (\eg Linux Auditd or CamFlow),
  which would negatively affect performance on the entire machine.

\subsection{Lightweight Ad-hoc Access Control}
\label{sec:usecase:ac}
While \system was designed primarily to provide secure auditing,
  it can also be used to implement simple access control policy
  within the scope of a container.\footnote{Policy conflict resolution across \texttt{cgroup}
  hierarchy is out of scope of this paper. We refer interested readers to \autoref{sec:rw}.}
We implemented a proof-of-concept to demonstrate \system's ability to 
  enforce access control policy.
Like in \autoref{sec:usecase:unicorn}, we consider a Kubernetes environment 
  and use the sidecar pattern to deploy access control policy at the \emph{pod} granularity.

\noindgras{Design \& Implementation.}
Using \system, we can easily achieve \emph{separation of concerns} in Kubernetes,
  such that each pod has its own set of security mechanisms and policies.
We deploy a sidecar alongside unmodified applications to constrain their behavior.
The sidecar runs a set of \system programs implementing the desired policy and 
  attach them to the root \texttt{cgroup} of the pod.
We associate security contexts to kernel objects through local storage
  and define an \ebpf map to store constraints applicable to those contexts.
When an LSM hook is triggered, information is retrieved from the map to
  determine whether or not an action is permitted.
Policy violation can be sent to userspace via an \ebpf ring buffer,
which can then be logged or reported to the user about the unexpected application behavior.
  
We take advantage of the \ebpf framework to optimize the sidecar
  at the time of its compilation based on the policy to be enforced.
For example, if the policy has no network access rules, 
  we do not build any rules enforcement program regarding network access.
In general, our access control mechanism generates a minimum set of programs 
  needed to enforce a given policy, thus reducing complexity and improving performance.

\noindgras{Policy Example.}
Taking inspiration from the Open Policy Agent~\cite{opa} and AppArmor~\cite{apparmor}, 
  we create a simple policy language.
A policy is expressed in JSON and parsed to generate a customized 
  sidecar application that can be attached to a pod.

\begin{lstlisting}[float=h, numbers=left, language=json, style=customC, caption={A simple policy example.}, label=listing:policy]
{
  "target": "/usr/bin/foo",
  "network": {"default": "deny", "allow_egress": [404,80]},
  "file": {"default": "r", "rw": ["/tmp/*"], "mr": ["/lib/ld-*.so*",
   "/lib/lib*.so*"]}
}
\end{lstlisting}

\autoref{listing:policy} shows an exemplar policy written in this language.
A \texttt{/usr/bin/foo} process is by default denied access to the network unless
  it is an outgoing connection through the \texttt{http} and \texttt{https} ports.
Similarly, the process is denied \texttt{write} or \texttt{execute} by default.
  However, it has read and write access to the \texttt{/tmp} directory and is able to 
  map system libraries.
This policy is inherited by any child process \texttt{fork}ed from the \texttt{/usr/bin/foo} process.

\subsection{\reviewchange{Discussion}}
We conclude this section by summarizing the advantages of using \system,
  as repeatedly demonstrated by the three use cases described above.

\noindgras{Performance.}
\system is the first reference-monitor-based audit system that allows audit rule configuration
  at compilation time, drastically minimizing run-time audit complexity 
  and improving overall performance.
In stark contrast, other audit mechanisms such as \camflow must evaluate
  complex audit rules at runtime to satisfy specific needs of different downstream applications.
For example, security tools such as SIGL~\cite{han2020sigl} typically analyze only a small 
  subset of host activity logs that an audit system like \camflow provides.
To monitor an application in a Kubernetes pod, Unicorn requires provenance data generated
  only by activity in the pod. 
In both cases, filtering is inevitable but it can sometimes become a performance bottleneck
  that is difficult to overcome.
To make matters worse, 
  as we have discussed in~\autoref{sec:policy:lsm}, run-time evaluation can have
  adverse and cumulative performance impact, making existing reference monitors
  undesirable to be even considered in practical settings.
Similarly, a given application may only enforce access control constraints on a subset of events.
Through compilation-time policy evaluation, 
  \system can minimize run-time cost by running only needed \ebpf programs.

In practice, this means that given an equivalent policy, 
  a solution built using \system significantly outperforms current solutions developed through
  the built-in LSM mechanism. 

\noindgras{Maintainability and Adoption.}
Maintaining out of tree LSMs requires significant effort and time investment.
As LSMs are built in the kernel, rigorous testing is essential to avoid crashes or introducing
  unintended security vulnerabilities.
This makes exploring new mechanisms difficult.
For similar reasons, it is rare for third parties to further develop on a custom kernel
  given the high risk of instability and vulnerability.
We further discuss maintainability concerns in \autoref{sec:discussion}.

\noindgras{Decentralized Deployment.} 
Standard LSM-based solutions are generally deployed system-wide and centralized,
 and must be managed by the host.
By contrast, each containerized environment (assuming proper privileges) can 
  deploy its own LSM mechanisms using \system without affecting the rest of the system.
Each guest environment can run not only different policies, 
  but also a completely different mechanism. 
Moreover, \system programs are only triggered within the \texttt{cgroup} 
  they are attached to, thus limiting data leakage across containers
  (see \autoref{sec:discussion} for further discussion on security).

\section{Understanding Policy Overhead}
\label{sec:lsm-bpf}
The run-time performance overhead of any always-on audit system is critical
  to its successful adoption.
While the overall performance is a function of a specific audit policy,
  which varies across different needs and use cases,
  the run-time cost of the underlying infrastructure can be reasonably analyzed,
  which we present in this section.
Our analysis focuses on two main components of \system, LSM and \ebpf;
the cost of running both together has not been widely studied,
  especially in the context of audit.
We also compare our approach with state-of-the-art reference-monitor-based
  auditing that requires kernel modification.

\subsection{\lsm overhead}
\label{sec:policy:lsm}
It is difficult to precisely measure LSM overhead~\cite{zhang2021analyzing}.
In general, there exist two sources of overhead when performing audit
  (or other policy enforcement) through LSM: \emph{hooking}
  and \emph{execution}.
Hooking refers to the cost of invoking a callback function associated
  with a specific LSM hook, which incurs roughly constant overhead.
Execution refers to the cost of running the callback function,
  which is dependent on the specific audit (or other policy enforcement) mechanism
  and can also vary by the (audit) policy itself.

It is sometimes mistakenly assumed that
  for a given system call and a given policy on the
  system call, the overhead introduced by a specific LSM module would be constant.
In reality, such an assumption is often an oversimplification.
Consider an \texttt{open} system call.
A number of LSM hooks, such as \texttt{file\_open} and \texttt{inode\_permission},
  are triggered when \texttt{open} is called.
If a new file is created because of \texttt{open},
  additional hooks such as \texttt{inode\_create} and \texttt{inode\_setattr} are called
  when the new file's underlying inode is being created and its attributes set.
Of particular interest in this example is the \texttt{inode\_permission} hook,
  which is called on each directory composing the path of the file to be opened,
  since \texttt{open} must have the permission to search for the file to be opened.

To audit an open-file event,
  it is important to record all the permission checks (including the ones
  on the directories) because it reveals file access patterns.
For example, in a security context, a failed \texttt{inode\_permission} check could
  indicate that a compromised application attempted to scan the file system to access sensitive data.
The overhead introduced by such an audit mechanism on this particular
  event is a function of path length.
For example, assume that invoking \texttt{file\_open}
  and \texttt{inode\_permission} and running their callback functions incur
  the same cost $C$.
  The total overhead of a file-open event on a path of length $N$ is
  $C \times (N+1)$.
Given two audit policies, $P_{A}$ and $P_{B}$, such that $C_{A}$ is one
  order of magnitude higher than $C_{B}$, the total overhead incurred by
  $P_{A}$ is in fact \emph{two} orders of magnitude higher than that by
  $P_{B}$ on a path of length 10.
The \texttt{open} system call is not the only one affected by such behavior;
  other system calls, such as \texttt{chmod}, \texttt{symlink}, \texttt{mmap},
  \texttt{stat}, and \texttt{execve} have similar patterns.
  
\begin{table}[t]
	\begin{tabularx}{\columnwidth}{l|X|c}
		System Call & Security Hooks                                                                      & Min Hook Calls       \\
		\hline
		\texttt{open}        & file\_open\textbf{+} \newline  inode\_create \newline  inode\_permission\textbf{*} \newline  inode\_post\_setxattr \newline  inode\_setattr & 1 + 1 $\times$ path depth \\
		\hline
		\texttt{read}        & file\_permission\textbf{+}                                                                    & 1                  \\
		\hline
		\texttt{write}       & file\_permission\textbf{+}                                                                    & 1                  \\
		\hline
		\texttt{execve}      & bprm\_check\textbf{+} \newline  bprm\_set\_creds\textbf{+} \newline  file\_open\textbf{+} \newline  inode\_permission\textbf{*} \newline  file\_permission\textbf{+}      & 4 + 1 $\times$ path depth
	\end{tabularx}
	\vspace{1mm}
	\caption{Summary of LSM hooks called on successful system calls.
		Some hooks are only triggered in a particular system state or with
		specific syscall parameters (\eg when creating a \emph{new} file
		on \texttt{open}). \textbf{+} indicates that hooks are always called,
		and \textbf{*} means hooks are called on every directory in a path.}
	\label{table:syscall}
\end{table}

Because this phenomenon can have a significant impact
  on the overall system performance,
we analyze the call graph associated with each system call (see \autoref{sec:usecase:provbpf}) to understand LSM hook invocation patterns.
We show the results for a few system calls in~\autoref{table:syscall} (note that for readability, we do not include hooks that are called when a system call fails/errs).

\subsection{\system overhead}

\system's sources of overhead are
  fundamentally the same as those of standard LSM security modules,
  \ie hooking and execution (\autoref{sec:policy:lsm}).
Therefore, if a standard security module and \system implement
  the same policy, they incur roughly the same total overhead,
  except that \system incurs some additional cost to traverse the \texttt{cgroup} hierarchy and to invoke the relevant \ebpf programs (\autoref{sec:namespacing}).

In practice, however,  there exists significant differences in policy overhead
  between a standard security module and \system;
  \system offers time-saving convenience and flexibility that a
  standard security module is unable to provide.
To run a customized in-kernel LSM module,
  the Linux kernel must be modified.
This requires thorough testing before the deployment of the custom kernel.
It is common for average users to shy away from the mere idea of deploying
  a kernel running heavily-customized code,
  especially one where said customized code interacts with the OS security framework.
To mitigate those issues, standard modules are typically designed to be general-purpose.
For example, an auditing module (\eg CamFlow~\cite{pasquier2017practical-socc2017})
  must be able to satisfy different auditing needs without requiring users to compile
  their own custom kernel.
To that end, the module must evaluate \emph{at runtime} an extensive audit policy
  to determine what information it should log.
As a concrete example, Bates~\etal~\cite{bates2015take} deploy
  policies to record events based on their security context as provided by SELinux.
For each object involved in a given event,
  the audit mechanism needs to retrieve its security ID and compare it with the specified policy.
While the policy is relatively simple,
  the cumulative effects (as discussed in~\autoref{sec:policy:lsm})
  on the policy have a significant impact on performance.

On the other hand,
  \system-based solutions take into account audit policy \emph{at compilation time},
  which significantly reduces run-time complexity and thus improves performance.
Moreover, since \system allows users to attach programs based on \texttt{cgroup}s,
  there is virtually no overhead imposed on applications running
  outside of the targeted \texttt{cgroup}.
This means, for example, that if a Kubernetes pod deploys a complex audit mechanism,
  the other pods on the system remain unaffected.

\section{Performance Evaluation}
\label{sec:evaluation}
In this section,
  we evaluate \system performance on a bare metal machine
  with 16GiB of RAM and an Intel i7 CPU.
In~\autoref{sec:evaluation:namespaced},
  we analyze the cost of hook invocation on \system.
Next, in~\autoref{sec:evaluation:provbpf},
  we explore the performance gain from using \system rather than a state-of-the-art
  monitoring system that modifies the Linux kernel,
  when performing exactly the same functionality.
\autoref{sec:availability} has more details on reproducing the results
  reported in this section.

\subsection{Overhead of Namespacing}
\label{sec:evaluation:namespaced}

\begin{figure}[t]
	\pgfplotsset{width=\columnwidth,compat=1.10}
\begin{tikzpicture}
	\centering
	\begin{axis}[
		ybar, axis on top,
		bar width=0.4cm,
		ymajorgrids, tick align=inside,
		major grid style={draw=white},
		ymin=0, ymax=20,
		axis x line*=bottom,
		axis y line*=right,
		y axis line style={opacity=0},
		tickwidth=0pt,
		enlarge x limits=true,
		legend style={
			legend columns=-1,
			/tikz/every even column/.append style={column sep=0.5cm}
		},
		ylabel={Time (relative)},
		symbolic x coords={
			socket,bind,listen,accept},
		xtick=data,
		]
		\addplot [draw=none, fill=blue!30] coordinates {
			(socket,1)
			(bind,0.74) 
			(listen,0.72)
			(accept,0.73) };
		\addplot [draw=none,fill=red!30] coordinates {
			(socket,9.77)
			(bind,9.2) 
			(listen,9.34)
			(accept,9.38) };
		\addplot [draw=none, fill=green!30] coordinates {
			(socket,15.77)
			(bind,15.88)
			(listen,15.85)
			(accept,15.27) };
		
		\legend{LSM, LSM-BPF, \system}
	\end{axis}
\end{tikzpicture}
	\caption{Overhead of the LSM, LSM-BPF, and \system invocation mechanisms.}
	\label{image:ftrace_hooks}
\end{figure}
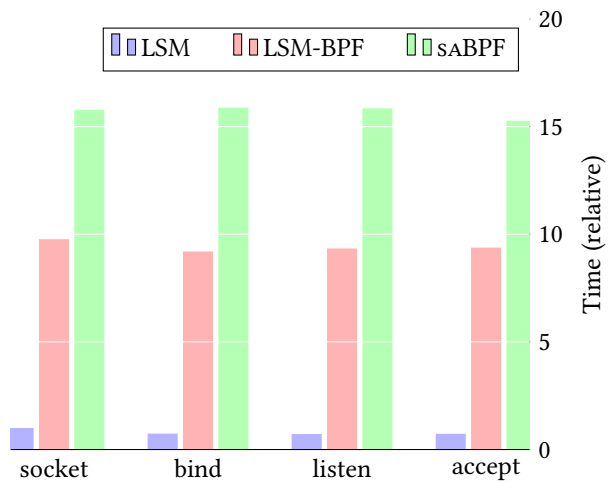

We compare the overhead associated with different mechanisms
  responsible for calling LSM hooks.
We are interested in the following three strategies:
1) the native LSM mechanism with built-in functions (LSM);
2) LSM-BPF that attaches an \ebpf function to an LSM hook (LSM-BPF);
and 3) \system that attaches \ebpf programs at the intersection of a \texttt{cgroup} and an LSM hook (\system).
We use \emph{ftrace}~\cite{ftracedoc} to perform the measurement.
To measure exclusively the cost of each calling mechanism,
  the function or program that is attached to each hook
  performs no operations and simply returns.
We capture the overhead of four common functions associated with a UNIX server:
\texttt{socket}, \texttt{bind}, \texttt{listen}, and \texttt{accept}.
The overhead is associated with the following four LSM hooks:
\texttt{security\_socket\_[create/bind/listen/accept]}.
The results are shown in~\autoref{image:ftrace_hooks}.
We report overhead \emph{relative} to a single baseline and
  focus on order-of-magnitude comparison, because ftrace (or any similar kernel instrumentation tool)
  can introduce additional overhead~\cite{ftracedoc}.
We use the overhead of native LSM on the \texttt{socket\_create} hook to normalize experimental results.

\begin{algorithm}[t]
	\SetAlgoLined
	{\color{blue}// disallow task core migration and preemption}\\
	migrate\_disable()\;
	rcu\_read\_lock()\;
	rc = run\_bpf\_programs()\;
	rcu\_read\_unlock()\;
	{\color{blue}// allow task core migration and preemption}\\
	migrate\_enable()\;
	\Return rc\;
	\caption{Execute an \ebpf program (simplified).}
	\label{algo:bpf}
\end{algorithm}

We see in~\autoref{image:ftrace_hooks} that the invocation overhead
  is nearly constant and independent of the system call.
It is roughly 10 and 15 times more costly with LSM-BPF and \system
  than with native LSM, respectively.
The built-in LSM simply finds the address of the LSM function in a hook table
  and calls the function.
The extra cost of LSM-BPF is related to the cost of invoking an \ebpf program.
We show the simplified logic to execute an \ebpf program in~\autoref{algo:bpf}.
While the overhead of executing the \ebpf program itself is relatively low
  (close to executing a native function),
  handling read-copy-update (RCU)~\cite{rcu} synchronization primitives and
  manipulating scheduler migration and preemption flags
  accounts for the majority of the overhead.
  
\begin{algorithm}[t]
	\SetAlgoLined
	hierarchy = get\_cgroup\_hierarchy(current\_task, hook\_reference)\;
	{\color{blue}// disallow task core migration and preemption}\\
	migrate\_disable()\;
	rcu\_read\_lock()\;
	\ForEach{cgroup in hierarchy}{
		rc = run\_bpf\_programs()\;
		\If{rc}{\Return rc\;}
	}
	rcu\_read\_unlock()\;
	{\color{blue}// allow task core migration and preemption}\\
	migrate\_enable()\;
	\Return rc\;
	\caption{Execute~an~\system~program~(simplified).}
	\label{algo:sabpf}
\end{algorithm}

As shown in~\autoref{algo:sabpf},
  \system follows a similar logic, except that it incurs additional overhead
  when retrieving and traversing the \texttt{cgroup} hierarchy.

However,  we emphasize that these relative overheads must be
  considered with respect to the cost of policy evaluation.
As a point of comparison, SELinux's policy evaluation cost of the \texttt{socket\_create} hook is $2,000$ times larger than the invocation cost of native LSM.
To better understand the actual cost of running \system and to contextualize its overhead,
  we perform both micro- and macro-benchmarks in the next section.

\subsection{Evaluating \capturesystem}
\label{sec:evaluation:provbpf}

To contextualize \system's performance with a realistic workload,
  we perform an evaluation of \capturesystem through micro- (\autoref{sec:evaluation:micro}) and macro-benchmarks (\autoref{sec:evaluation:macro}).
We demonstrate that \capturesystem outperforms
  the state-of-the-art whole-system provenance solution \camflow~\cite{pasquier2017practical-socc2017} and incurs minimal performance overhead.

We choose standard benchmarks such as \texttt{lmbench}, so that \system
can be meaningfully compared with prior and future work.
We run each benchmark on three different kernel configurations.
The \emph{vanilla} configuration runs on the unmodified mainline Linux
kernel v5.11.2, which serves as our baseline.
The \emph{CamFlow} configuration uses the same kernel but additionally
instrumented with \camflow kernel patches (v0.7.2)~\cite{camflow_web}.
Finally, the \capturesystem workload corresponds to the same Linux kernel
but running with our \ebpf-based provenance capture mechanism \capturesystem.
We also ensure that \capturesystem's and \camflow's configurations are equivalent.

\subsubsection{Microbenchmark}
\label{sec:evaluation:micro}

\begin{table}[t]
	\resizebox{\columnwidth}{!}{
		\begin{tabular}{ l | c c c c c}
			Test Type & vanilla & CamFlow & Overhead & \capturesystem & Overhead \\
			\hline
			\multicolumn{6}{c}{Process tests (in $\mu s$, the smaller the better)}\\
			\hline
			NULL call			& 0.30		& 0.32 		& 0\%  		& 0.29     	& 0\%\\
			NULL I/O			& 0.39		& 0.75 		& 92\%  	& 0.54     	& 38\%\\
			stat 				& 1.04		& 3.77 		& 263\%  	& 1.40     	& 35\%\\
			fstat 				& 0.52		& 1.40 		& 169\%  	& 0.66     	& 28\%\\
			open/close file 	& 1.80		& 5.89 		& 227\%  	& 2.62	  	& 46\%\\
			read file 	& 0.40		& 0.73 		& 84\%  	& 0.56	  	& 42\%\\
			write file 	& 0.36		& 0.70 		& 92\%  	& 0.53	  	& 53\%\\
			fork process		& 295.55 	& 344.15 	& 13\%  	& 317.78	& 8\%\\
			\hline
			\multicolumn{6}{c}{File and memory latency (in $\mu s$, the smaller the better)}\\
			\hline
			file create (0k)	& 10.31 	& 21.20	& 106\%  & 13.10  & 27\%\\
			file delete (0k)	& 11.25		& 23.35	& 108\%  & 12.80  & 14\%\\
			file create (10k)	& 16.55		& 40.75	& 146\%  & 20.65  & 25\%\\
			file delete (10k)	& 13.45		& 30.20	& 125\%  & 15.55  & 16\%\\
			pipe latency		& 6.06		& 10.45 & 72\% 	 & 6.55   & 8\%\\
			AF\_UNIX latency	& 6.60		& 16.43 & 149\%	 & 9.72   & 47\%\\
		\end{tabular}
	}
  \vspace{1mm}
	\caption{\texttt{lmbench} results.}
	\label{table:eval:lmbench}
\end{table}

We use \texttt{lmbench}~\cite{mcvoy1996lmbench} to measure
  \capturesystem's performance overhead on raw system calls,
as reported in \autoref{table:eval:lmbench}.
We show only a relevant subset of performance metrics
due to space constraints, but the complete results are available online
(see \autoref{sec:availability}).

The overhead of \capturesystem, when compared to the vanilla kernel, is relatively low.
In addition to the overhead introduced by the invocation mechanism,
  \capturesystem also incurs the cost of building the provenance graph elements and sending them to the user-space program.
It outperforms \camflow as it is significantly streamlined.
Indeed, \camflow uses a complex set of capture policies to allow
  users to tailor data capture to their specific needs~\cite{pasquier2017practical-socc2017}.
Evaluating the policy at runtime can be relatively costly,
  especially since the effects can be cumulative (\autoref{sec:lsm-bpf}).
In the case of \capturesystem, policy evaluation is performed at compilation time,
  so that the compiled code only captures the desired events, thus significantly reducing overhead given equivalent policies.

\subsubsection{Macrobenchmark}
\label{sec:evaluation:macro}

We present two sets of macrobenchmarks.
The first set, as shown in~\autoref{table:eval:macrobenchmark},
  measures the performance impact on a single machine
  when \emph{unpack}ing and \emph{build}ing the kernel and
  running the \emph{Postmark} benchmark\reviewchange{~\cite{katcher1997postmark}}.
These are the common benchmarks used in prior provenance
  literature ever since Muniswamy-Reddy~\etal~\cite{muniswamy2006provenance}
  introduced the concept of system provenance.
\autoref{table:eval:cloudybench} shows the results of the
  second set of benchmarks focusing on a set of applications
  typically used to build web applications.
\reviewchange{These benchmarks are not intended 
to cover every possible scenario,
but rather to provide meaningful points of comparison.	
We rely on the Phoronix Test Suite~\cite{phoronix} to perform these benchmarks.
Details on benchmark parameters and settings are available in our repository, see \autoref{sec:availability}.}
	
\begin{table}[t]
	\resizebox{\columnwidth}{!}{
		\begin{tabular}{ l | c c c c c}
			Test Type & vanilla &  CamFlow & Overhead & \capturesystem & Overhead \\
			\hline
			\multicolumn{6}{c}{Execution time (in seconds, the smaller the better)}\\
			\hline
			unpack 				& 6.52   & 7.70   & 18\% & 6.59   & 1\% \\
			build 				& 194.26 & 232.01 & 19\% & 203.70 & 5\% \\
			\hline
			\multicolumn{6}{c}{4kB to 1MB file, 10 subdirectories,4k5 simultaneous transactions, 1M5 transactions}\\
			\hline
			postmark 			& 79.50 & 113.00 & 42\% & 92.50 & 16\% \\
			\hline
		\end{tabular}
	}
  \vspace{1mm}
	\caption{Macrobenchmark results.}
	\label{table:eval:macrobenchmark}
\end{table}

\begin{table}[t]
	\begin{center}
		\resizebox{\columnwidth}{!}{
			\begin{tabular}{ l | c c c c c}
				Test Type & vanilla & CamFlow & Overhead & \capturesystem & Overhead \\
				\hline
				\multicolumn{6}{c}{Request/Operation per second (the higher the better)}\\
				\hline
				apache httpd			& 14645 	& 10682   & 27\% & 13487 	& 8\% \\
				redis (LPOP)		   	& 2105221 	& 1780868 & 15\% & 1894961 	& 10\% \\
				redis (SADD)			& 2073489	& 1721367 & 17\% & 1854162 	& 11\% \\
				redis (LPUSH)			& 1630446 	& 1401497 & 14\% & 1510000 	& 7\% \\
				redis (GET)				& 2360694 	& 1928276 & 18\% & 2102901 	& 11\% \\
				redis (SET)				& 1873359 	& 1569507 & 16\% & 1690189 	& 10\% \\
				memcache (ADD)			& 44122 	& 30444   & 31\% & 41362 	& 6\% \\
				memcache (GET)			& 67895 	& 41363   & 39\% & 62167 	& 8\% \\
				memcache (SET)			& 44460 	& 30346   & 32\% & 41355 	& 7\% \\
				memcache (APPEND)	 	& 46730 	& 31157 	& 33\% & 43215 	& 8\% \\
				memcache (DELETE)		& 67761 	& 40735   & 40\% & 61755 	& 9\% \\
				php						& 690725 	& 613296  & 11\% & 709476 	& 0\% \\
				\hline
				\multicolumn{6}{c}{Execution time (in ms, the lower the better)}\\
				\hline
				pybench					& 1246 		& 1298 & 4\% & 1196 & 0\% \\
				\hline
			\end{tabular}
		}
	\end{center}
  \vspace{1mm}
	\caption{Extended macrobenchmark results.}
	\label{table:eval:cloudybench}
\end{table}

From the first set of benchmarks (\autoref{table:eval:macrobenchmark}),
  we see that \capturesystem introduces between 1\% and 16\% overhead.
Unpack and build workloads are computation heavy,
  and most of the execution time is spent in userspace.
On the other hand,  postmark spends a more significant portion of its execution time
  in system call code.
As \capturesystem only adds overhead when system calls are executed,
  it unsurprisingly performs worse in the Postmark benchmark.
In the second set of benchmarks (\autoref{table:eval:cloudybench}),
  we evaluate the impact of \capturesystem on applications that are often deployed
  through containers.
\capturesystem's overhead is between 0\% and 11\%.
In all scenarios, \capturesystem outperforms \camflow.

We also note that \capturesystem results are in the same order of magnitude
  as similar whole-system provenance capture solutions
  such as Hi-Fi~\cite{pohly2012hi} and LPM~\cite{bates2015trustworthy}.
We are not able to provide direct comparison with these solutions
  since they were implemented for extremely outdated kernels
  (release 2.6.32 from 2009 for LPM~\cite{lpmsrc} and release 3.2.0 from 2011 for Hi-Fi~\cite{hifisrc});
  internal kernel changes make it practically impossible for us to port
  them to a modern kernel release.

\section{Discussion}
\label{sec:discussion}
\noindgras{Security.}
We are aware of a number of security issues with \ebpf
  (\eg CVE~\cite{CVE-2020-8835} and CVE~\cite{CVE-2021-29154}).
In many known attack scenarios related to \ebpf,
  an attacker exploits the \ebpf verifier to make illegal modifications of kernel data structures,
  \eg to perform privilege escalation~\cite{CVE-2020-8835}.
One clear solution is to improve the verification of \ebpf programs~\cite{gershuni2019simple,nelson2020specification}.
While this is an important problem worthy of investigation,
  it is orthogonal to \system and therefore out of scope for this paper.
We note that, to the best of our knowledge, \system does not
  introduce new attack vectors and that
  any improvement to \ebpf security will
  benefit \system.

\noindgras{Layering.}
In this work, we focus on capturing kernel-level audit data that
  describes low-level system interactions.
However, to fully understand application behavior,
  it is often useful to analyze audit information from multiple sources,
  preferably from different layers of abstraction.
For example, layering both low-level system traces and higher-level
  application traces can often facilitate attack investigation
  by enabling forensic experts to
  identify, in an iterative fashion, an attack point of entry~\cite{lee2013high}.
The application of such techniques is beyond the scope of this paper,
  but \system and any application built atop can be
  seamlessly integrated with existing layering techniques.

\begin{table}[]
	\begin{tabularx}{\columnwidth}{l| c | c | c}
		Date & Release & Long Term Support & Changes\\
		\hline
		April 2021 & 5.12 & No & 4\\
		February 2021 & 5.11 & No & 3\\
		December 2020 & 5.10 & Yes & 2\\
		October 2020 & 5.9 & No & 0\\
		August 2020 & 5.8 & No & 4\\
		May 2020 & 5.7 & No & 0\\
		March 2020 & 5.6 & No & 0\\
		January 2020 & 5.5 & No & 5\\
		November 2019 & 5.4 & Yes & 1\\
	\end{tabularx}
  \vspace{1mm}
	\caption{Changes made to the LSM ABI in terms of the number of interface function modified (including name changes, parameter modifications, and additions and deletions) since the latest release. We note that there is a total of 236 LSM hooks as of release 5.12.}
	\label{table:abichange}
\end{table}

\noindgras{Maintainability.}
One of the key advantages in building audit tools through \ebpf
  and by extension \system is that they can be heavily customized to
  fulfill the needs of the user.
As we previously pointed out, maintaining bespoke
  built-in audit tools requires the developers to, at a minimum,
  1) maintain a custom kernel,
  2) prepare a custom OS distribution, and
  3) perform extensive testing before actual deployment.
This burden is greatly alleviated using our proposed solution.
The audit mechanism is neatly separated from the OS
  and can be built and tested independently.
Furthermore, with BTF and CO-RE\cite{CORE-2020},
  any solution built with \system does not need
  to be built against a specific version of the kernel;
  it only needs to be rebuilt (and updated) when the kernel's internal LSM ABI changes, which is rare (\autoref{table:abichange}).
We note that a number of popular distributions ship only Long Term Support kernel versions,
  which further simplifies maintenance.

\section{Related Work}
\label{sec:rw}
\system is designed mostly for monitoring containers in the cloud and uses
  two major technologies, \ebpf and LSM.
We discuss related work in these areas.

\noindgras{\ebpf-based Security.}
In system security, 
  one of the well-known \ebpf-enabled applications is \texttt{seccomp-bpf},
  which filters system calls available for user-space
  applications to reduce kernel attack surface~\cite{edge2015seccomp}.
\texttt{seccomp} filters use
  BPF programs to decide, based on the system call number and
  arguments, whether a given call is allowed or not.
A more recent application of \ebpf is LBM~\cite{tian2019lbm},
  which protects the Linux kernel from malicious peripherals such as USB,
   Bluetooth, and NFC\reviewchange{.}
LBM places interposition hooks, through the implementation of 
  new \ebpf program types,  right beneath a peripheral's protocol stack and
  above the peripheral's controller driver, so that it can guarantee
  that \ebpf programs can filter all inputs from the device and all outputs
  from the host. 
LBM introduces a new filter language for peripherals to enforce 
  programmable security policies.

LSM-BPF is still a nascent \ebpf extension, emerging from
  Kernel Runtime Security Instrumentation (KRSI)~\cite{krsi_lwn}.
KRSI enables privileged users to
  dynamically update MAC and audit policies
  based on the state of the computing environment.
\texttt{bpfbox}~\cite{findlay2020bpfbox} uses LSM-BPF 
  to create process sandboxes through a flexible policy language.
BPFContain~\cite{findlay2021bpfcontain} uses LSM-BPF to enforce
  system-wide policy to control container IPC,  and file and network access.
They both leverage \ebpf because it is easier to maintain and further
  develop \ebpf programs than to use out-of-tree LSMs for their particular 
  needs.
\system is orthogonal to these systems and focuses on leveraging \ebpf 
  at the intersection of \texttt{cgroup} and LSM hooks, 
  mainly but not exclusively for secure auditing.

\noindgras{Monitoring Containers.}
There are a number of widely available solutions, such as Cilium~\cite{cilium}, Grafana~\cite{grafana}, and Nagios~\cite{nagios}, that monitor containers,
  but they focus primarily on performance and/or network traffic monitoring.
We design \system not to compete with these solutions, but to allow their functionality to be 
  extended to secure auditing.
Indeed, while these frameworks provide a wealth of information, 
  their capture methodology does not provide strong enough guarantees 
  in the presence of an attacker.
Furthermore, \system enables the implementation of decentralized solutions that
  need not be managed by the host platform.

\noindgras{LSM.}
The LSM framework~\cite{morris2002linux} was introduced nearly two decades ago
  to Linux for Mandatory Access Control (MAC).
Two of the most popular applications of LSM are AppArmor~\cite{apparmor} and SELinux~\cite{smalley2001implementing}.
Over the years, the LSM framework has seen its usage 
  extended to implementing mechanisms such as 
  the Linux Integrity Measurement Architecture~\cite{sailer2004design},
  which enables hardware-based integrity attestation,
  and loadpin~\cite{loadpin},
  which was developed to restrict the origin of kernel-loaded code to 
  read-only devices in ChromeOS.
LSM has also been used to implement secure auditing as previously mentioned~\cite{pohly2012hi,bates2015trustworthy,pasquier2017practical-socc2017}.
These work is orthogonal to \system; %
instead, \system is closely related to prior work that attempted
  to allow namespacing and stacking of LSM modules~\cite{sun2018security,namespacing2017}.
They focused on enabling containers to define their own security policy
  within a system-wide MAC scheme.
For example, Sun \etal~\cite{sun2018security} make AppArmor namespace-aware
  so that each individual container can have its own policy to be enforced by the host.
This is a non-trivial task involving conflict resolution alongside the security namespace hierarchy.
\system expends on such ideas by allowing containers to provide not only their own policy, 
  but also their own totally separate mechanisms.

\section{Conclusion}
\label{sec:conclusion}
We present \system,
  a lightweight system-level auditing framework
  for container-based cloud environments.
\system is built upon the widely-used \ebpf framework. 
It is simple to use
  and allows individual containers to deploy -- in a decentralized manner -- 
  secure auditing tools.
These tools in turn enable users to implement
  a wide range of security solutions on container-oriented cloud platforms,
  such as intrusion detection systems for individual containers.
Using \system, we were able to re-implement a state-of-the-art audit system
  that provides exactly the same functionality while significantly improving performance.
We open-source \system, welcoming the community to develop and deploy
  toolsets that leverage our system,
  in hopes of further discovering the potentials of \system.

\appendix

\section{Availability}
\label{sec:availability}
Our implementation  (\autoref{sec:extending}) and instructions to reproduce the results presented (\autoref{sec:evaluation}) are available at \url{https://github.com/saBPF-project}.

\bibliographystyle{ACM-Reference-Format}
\bibliography{biblio}


\begin{thebibliography}{68}


\ifx \showCODEN    \undefined \def \showCODEN     #1{\unskip}     \fi
\ifx \showDOI      \undefined \def \showDOI       #1{#1}\fi
\ifx \showISBNx    \undefined \def \showISBNx     #1{\unskip}     \fi
\ifx \showISBNxiii \undefined \def \showISBNxiii  #1{\unskip}     \fi
\ifx \showISSN     \undefined \def \showISSN      #1{\unskip}     \fi
\ifx \showLCCN     \undefined \def \showLCCN      #1{\unskip}     \fi
\ifx \shownote     \undefined \def \shownote      #1{#1}          \fi
\ifx \showarticletitle \undefined \def \showarticletitle #1{#1}   \fi
\ifx \showURL      \undefined \def \showURL       {\relax}        \fi
\providecommand\bibfield[2]{#2}
\providecommand\bibinfo[2]{#2}
\providecommand\natexlab[1]{#1}
\providecommand\showeprint[2][]{arXiv:#2}

\bibitem[\protect\citeauthoryear{??}{kaf}{[n.d.]}]%
        {kafka}
 \bibinfo{year}{[n.d.]}\natexlab{}.
\newblock \bibinfo{title}{{Apche Kafka}}.
\newblock \bibinfo{howpublished}{online \reviewchange{(accessed \today)}}.
\newblock
\newblock
\shownote{\url{https://kafka.apache.org/}.}


\bibitem[\protect\citeauthoryear{??}{app}{[n.d.]}]%
        {apparmor}
 \bibinfo{year}{[n.d.]}\natexlab{}.
\newblock \bibinfo{title}{{AppArmor}}.
\newblock \bibinfo{howpublished}{online \reviewchange{(accessed \today)}}.
\newblock
\newblock
\shownote{\url{https://apparmor.net/}.}


\bibitem[\protect\citeauthoryear{??}{bpf}{[n.d.]}]%
        {bpf-ringbuffer}
 \bibinfo{year}{[n.d.]}\natexlab{}.
\newblock \bibinfo{title}{{BPF ring buffer}}.
\newblock \bibinfo{howpublished}{online \reviewchange{(accessed \today)}}.
\newblock
\newblock
\shownote{\url{https://www.kernel.org/doc/html/latest/bpf/ringbuf.html}.}


\bibitem[\protect\citeauthoryear{??}{cam}{[n.d.]}]%
        {camflow_web}
 \bibinfo{year}{[n.d.]}\natexlab{}.
\newblock \bibinfo{title}{{CamFlow}}.
\newblock \bibinfo{howpublished}{online \reviewchange{(accessed \today)}}.
\newblock
\newblock
\shownote{\url{https://camflow.org/}.}


\bibitem[\protect\citeauthoryear{??}{cil}{[n.d.]}]%
        {cilium}
 \bibinfo{year}{[n.d.]}\natexlab{}.
\newblock \bibinfo{title}{{Cilium}}.
\newblock \bibinfo{howpublished}{online \reviewchange{(accessed \today)}}.
\newblock
\newblock
\shownote{\url{https://cilium.io/}.}


\bibitem[\protect\citeauthoryear{??}{CVE}{[n.d.]a}]%
        {CVE-2020-8835}
 \bibinfo{year}{[n.d.]}\natexlab{a}.
\newblock \bibinfo{title}{{CVE-2020-8835}}.
\newblock \bibinfo{howpublished}{online \reviewchange{(accessed \today)}}.
\newblock
\newblock
\shownote{\url{https://cve.mitre.org/cgi-bin/cvename.cgi?name=CVE-2020-8835}.}


\bibitem[\protect\citeauthoryear{??}{CVE}{[n.d.]b}]%
        {CVE-2021-29154}
 \bibinfo{year}{[n.d.]}\natexlab{b}.
\newblock \bibinfo{title}{{CVE-2021-29154}}.
\newblock \bibinfo{howpublished}{online \reviewchange{(accessed \today)}}.
\newblock
\newblock
\shownote{\url{https://cve.mitre.org/cgi-bin/cvename.cgi?name=CVE-2021-29154}.}


\bibitem[\protect\citeauthoryear{??}{ebp}{[n.d.]}]%
        {ebpf_io}
 \bibinfo{year}{[n.d.]}\natexlab{}.
\newblock \bibinfo{title}{{eBPF}}.
\newblock \bibinfo{howpublished}{online \reviewchange{(accessed \today)}}.
\newblock
\newblock
\shownote{\url{https://ebpf.io/}.}


\bibitem[\protect\citeauthoryear{??}{ftr}{[n.d.]}]%
        {ftracedoc}
 \bibinfo{year}{[n.d.]}\natexlab{}.
\newblock \bibinfo{title}{{ftrace documentation}}.
\newblock \bibinfo{howpublished}{online \reviewchange{(accessed \today)}}.
\newblock
\newblock
\shownote{\url{https://www.kernel.org/doc/html/v4.17/trace/ftrace.html}.}


\bibitem[\protect\citeauthoryear{??}{gra}{[n.d.]}]%
        {grafana}
 \bibinfo{year}{[n.d.]}\natexlab{}.
\newblock \bibinfo{title}{{Grafana}}.
\newblock \bibinfo{howpublished}{online \reviewchange{(accessed \today)}}.
\newblock
\newblock
\shownote{\url{https://grafana.com/}.}


\bibitem[\protect\citeauthoryear{??}{nag}{[n.d.]}]%
        {nagios}
 \bibinfo{year}{[n.d.]}\natexlab{}.
\newblock \bibinfo{title}{{Nagios}}.
\newblock \bibinfo{howpublished}{online \reviewchange{(accessed \today)}}.
\newblock
\newblock
\shownote{\url{https://www.nagios.org/}.}


\bibitem[\protect\citeauthoryear{??}{opa}{[n.d.]}]%
        {opa}
 \bibinfo{year}{[n.d.]}\natexlab{}.
\newblock \bibinfo{title}{{Open Policy Agent}}.
\newblock \bibinfo{howpublished}{online \reviewchange{(accessed \today)}}.
\newblock
\newblock
\shownote{\url{https://www.openpolicyagent.org/}.}


\bibitem[\protect\citeauthoryear{??}{pho}{[n.d.]}]%
        {phoronix}
 \bibinfo{year}{[n.d.]}\natexlab{}.
\newblock \bibinfo{title}{{Phoronix test suite}}.
\newblock \bibinfo{howpublished}{online \reviewchange{(accessed \today)}}.
\newblock
\newblock
\shownote{\url{https://www.phoronix-test-suite.com/}.}


\bibitem[\protect\citeauthoryear{??}{rcu}{2021}]%
        {rcu}
 \bibinfo{year}{2021}\natexlab{}.
\newblock \bibinfo{title}{{RCU}}.
\newblock \bibinfo{howpublished}{online \reviewchange{(accessed \today)}}.
\newblock
\newblock
\shownote{\url{https://www.kernel.org/doc/Documentation/RCU/whatisRCU.txt}.}


\bibitem[\protect\citeauthoryear{Bates, Butler, and Moyer}{Bates
  et~al\mbox{.}}{2015a}]%
        {bates2015take}
\bibfield{author}{\bibinfo{person}{Adam Bates}, \bibinfo{person}{Kevin~RB
  Butler}, {and} \bibinfo{person}{Thomas Moyer}.}
  \bibinfo{year}{2015}\natexlab{a}.
\newblock \showarticletitle{{Take only what you need: leveraging mandatory
  access control policy to reduce provenance storage costs}}. In
  \bibinfo{booktitle}{\emph{Workshop on the Theory and Practice of Provenance
  (TaPP 15)}}. USENIX.
\newblock


\bibitem[\protect\citeauthoryear{Bates, Tian, Butler, and Moyer}{Bates
  et~al\mbox{.}}{[n.d.]}]%
        {lpmsrc}
\bibfield{author}{\bibinfo{person}{Adam Bates}, \bibinfo{person}{Dave~Jing
  Tian}, \bibinfo{person}{Kevin~RB Butler}, {and} \bibinfo{person}{Thomas
  Moyer}.} \bibinfo{year}{[n.d.]}\natexlab{}.
\newblock \bibinfo{title}{{LPM source code}}.
\newblock \bibinfo{howpublished}{online \reviewchange{(accessed \today)}}.
\newblock
\newblock
\shownote{\url{https://bitbucket.org/uf_sensei/redhat-linux-provenance-release/}.}


\bibitem[\protect\citeauthoryear{Bates, Tian, Butler, and Moyer}{Bates
  et~al\mbox{.}}{2015b}]%
        {bates2015trustworthy}
\bibfield{author}{\bibinfo{person}{Adam Bates}, \bibinfo{person}{Dave~Jing
  Tian}, \bibinfo{person}{Kevin~RB Butler}, {and} \bibinfo{person}{Thomas
  Moyer}.} \bibinfo{year}{2015}\natexlab{b}.
\newblock \showarticletitle{{Trustworthy whole-system provenance for the linux
  kernel}}. In \bibinfo{booktitle}{\emph{Security Symposium}}. USENIX,
  \bibinfo{pages}{319--334}.
\newblock


\bibitem[\protect\citeauthoryear{Belhajjame, B’Far, Cheney, Coppens,
  Cresswell, Gil, Groth, Klyne, Lebo, McCusker, Miles, Myers, and
  Sahoo}{Belhajjame et~al\mbox{.}}{2013}]%
        {w3cprovdm}
\bibfield{author}{\bibinfo{person}{Khalid Belhajjame}, \bibinfo{person}{Reza
  B’Far}, \bibinfo{person}{James Cheney}, \bibinfo{person}{Sam Coppens},
  \bibinfo{person}{Stephen Cresswell}, \bibinfo{person}{Yolanda Gil},
  \bibinfo{person}{Paul Groth}, \bibinfo{person}{Graham Klyne},
  \bibinfo{person}{Timothy Lebo}, \bibinfo{person}{Jim McCusker},
  \bibinfo{person}{Simon Miles}, \bibinfo{person}{James Myers}, {and}
  \bibinfo{person}{Satya Sahoo}.} \bibinfo{year}{2013}\natexlab{}.
\newblock \bibinfo{booktitle}{\emph{{PROV-DM: The PROV Data Model}}}.
\newblock \bibinfo{type}{{T}echnical {R}eport}. \bibinfo{institution}{W3C}.
\newblock


\bibitem[\protect\citeauthoryear{Chan, Cheney, Bhatotia, Pasquier, Gehani,
  Irshad, Carata, and Seltzer}{Chan et~al\mbox{.}}{2019}]%
        {chan2019provmark}
\bibfield{author}{\bibinfo{person}{Sheung~Chi Chan}, \bibinfo{person}{James
  Cheney}, \bibinfo{person}{Pramod Bhatotia}, \bibinfo{person}{Thomas
  Pasquier}, \bibinfo{person}{Ashish Gehani}, \bibinfo{person}{Hassaan Irshad},
  \bibinfo{person}{Lucian Carata}, {and} \bibinfo{person}{Margo Seltzer}.}
  \bibinfo{year}{2019}\natexlab{}.
\newblock \showarticletitle{{ProvMark: a provenance expressiveness benchmarking
  system}}. In \bibinfo{booktitle}{\emph{International Middleware Conference}}.
  ACM/IFIP, \bibinfo{pages}{268--279}.
\newblock


\bibitem[\protect\citeauthoryear{Corbet}{Corbet}{2016}]%
        {loadpin}
\bibfield{author}{\bibinfo{person}{Jonathan Corbet}.}
  \bibinfo{year}{2016}\natexlab{}.
\newblock \bibinfo{title}{{LoadPin}}.
\newblock \bibinfo{howpublished}{online \reviewchange{(accessed \today)}}.
\newblock
\newblock
\shownote{\url{https://lwn.net/Articles/682302/}.}


\bibitem[\protect\citeauthoryear{Edge}{Edge}{2015}]%
        {edge2015seccomp}
\bibfield{author}{\bibinfo{person}{Jake Edge}.}
  \bibinfo{year}{2015}\natexlab{}.
\newblock \showarticletitle{A seccomp overview}.
\newblock \bibinfo{journal}{\emph{Linux Weekly News}} (\bibinfo{year}{2015}).
\newblock


\bibitem[\protect\citeauthoryear{Edwards, Jaeger, and Zhang}{Edwards
  et~al\mbox{.}}{2002}]%
        {edwards2002runtime}
\bibfield{author}{\bibinfo{person}{Antony Edwards}, \bibinfo{person}{Trent
  Jaeger}, {and} \bibinfo{person}{Xiaolan Zhang}.}
  \bibinfo{year}{2002}\natexlab{}.
\newblock \showarticletitle{{Runtime verification of authorization hook
  placement for the Linux security modules framework}}. In
  \bibinfo{booktitle}{\emph{Conference on Computer and Communications Security
  (CCS'02)}}. ACM, \bibinfo{pages}{225--234}.
\newblock


\bibitem[\protect\citeauthoryear{Findlay, Barrera, and Somayaji}{Findlay
  et~al\mbox{.}}{2021}]%
        {findlay2021bpfcontain}
\bibfield{author}{\bibinfo{person}{William Findlay}, \bibinfo{person}{David
  Barrera}, {and} \bibinfo{person}{Anil Somayaji}.}
  \bibinfo{year}{2021}\natexlab{}.
\newblock \showarticletitle{{BPFContain: Fixing the Soft Underbelly of
  Container Security}}.
\newblock \bibinfo{journal}{\emph{arXiv}} (\bibinfo{year}{2021}).
\newblock


\bibitem[\protect\citeauthoryear{Findlay, Somayaji, and Barrera}{Findlay
  et~al\mbox{.}}{2020}]%
        {findlay2020bpfbox}
\bibfield{author}{\bibinfo{person}{William Findlay}, \bibinfo{person}{Anil
  Somayaji}, {and} \bibinfo{person}{David Barrera}.}
  \bibinfo{year}{2020}\natexlab{}.
\newblock \showarticletitle{{bpfbox: Simple Precise Process Confinement with
  eBPF}}. In \bibinfo{booktitle}{\emph{Cloud Computing Security Workshop
  (CCSW)}}. ACM, \bibinfo{pages}{91--103}.
\newblock


\bibitem[\protect\citeauthoryear{Gao, Gu, Kayaalp, Pendarakis, and Wang}{Gao
  et~al\mbox{.}}{2017}]%
        {gao2017containerleaks}
\bibfield{author}{\bibinfo{person}{Xing Gao}, \bibinfo{person}{Zhongshu Gu},
  \bibinfo{person}{Mehmet Kayaalp}, \bibinfo{person}{Dimitrios Pendarakis},
  {and} \bibinfo{person}{Haining Wang}.} \bibinfo{year}{2017}\natexlab{}.
\newblock \showarticletitle{{ContainerLeaks: Emerging security threats of
  information leakages in container clouds}}. In
  \bibinfo{booktitle}{\emph{International Conference on Dependable Systems and
  Networks (DSN'17)}}. IEEE/IFIP, \bibinfo{pages}{237--248}.
\newblock


\bibitem[\protect\citeauthoryear{Gehani and Tariq}{Gehani and Tariq}{2012}]%
        {gehani2012spade}
\bibfield{author}{\bibinfo{person}{Ashish Gehani} {and} \bibinfo{person}{Dawood
  Tariq}.} \bibinfo{year}{2012}\natexlab{}.
\newblock \showarticletitle{{SPADE: Support for Provenance Auditing in
  Distributed Environments}}. In \bibinfo{booktitle}{\emph{International
  Middleware Conference}}. \bibinfo{publisher}{Springer-Verlag},
  \bibinfo{pages}{101--120}.
\newblock


\bibitem[\protect\citeauthoryear{Georget, Jaume, Tronel, Piolle, and
  Tong}{Georget et~al\mbox{.}}{2017}]%
        {georget2017verifying}
\bibfield{author}{\bibinfo{person}{Laurent Georget}, \bibinfo{person}{Mathieu
  Jaume}, \bibinfo{person}{Fr{\'e}d{\'e}ric Tronel}, \bibinfo{person}{Guillaume
  Piolle}, {and} \bibinfo{person}{Val{\'e}rie Viet~Triem Tong}.}
  \bibinfo{year}{2017}\natexlab{}.
\newblock \showarticletitle{{Verifying the reliability of operating
  system-level information flow control systems in linux}}. In
  \bibinfo{booktitle}{\emph{International FME Workshop on Formal Methods in
  Software Engineering (FormaliSE)}}. IEEE, \bibinfo{pages}{10--16}.
\newblock


\bibitem[\protect\citeauthoryear{Gershuni, Amit, Gurfinkel, Narodytska, Navas,
  Rinetzky, Ryzhyk, and Sagiv}{Gershuni et~al\mbox{.}}{2019}]%
        {gershuni2019simple}
\bibfield{author}{\bibinfo{person}{Elazar Gershuni}, \bibinfo{person}{Nadav
  Amit}, \bibinfo{person}{Arie Gurfinkel}, \bibinfo{person}{Nina Narodytska},
  \bibinfo{person}{Jorge~A Navas}, \bibinfo{person}{Noam Rinetzky},
  \bibinfo{person}{Leonid Ryzhyk}, {and} \bibinfo{person}{Mooly Sagiv}.}
  \bibinfo{year}{2019}\natexlab{}.
\newblock \showarticletitle{{Simple and precise static analysis of untrusted
  linux kernel extensions}}. In \bibinfo{booktitle}{\emph{Conference on
  Programming Language Design and Implementation (PLDI'19)}}. ACM,
  \bibinfo{pages}{1069--1084}.
\newblock


\bibitem[\protect\citeauthoryear{Han, Pasquier, Bates, Mickens, and
  Seltzer}{Han et~al\mbox{.}}{2020}]%
        {han2020ndss}
\bibfield{author}{\bibinfo{person}{Xueyuan Han}, \bibinfo{person}{Thomas
  Pasquier}, \bibinfo{person}{Adam Bates}, \bibinfo{person}{James Mickens},
  {and} \bibinfo{person}{Margo Seltzer}.} \bibinfo{year}{2020}\natexlab{}.
\newblock \showarticletitle{{UNICORN: Runtime Provenance-based Detector for
  Advanced Persistent Threats}}. In \bibinfo{booktitle}{\emph{Network and
  Distributed System Security Symposium (NDSS'20)}}. Internet Society.
\newblock


\bibitem[\protect\citeauthoryear{Han, Pasquier, Ranjan, Goldstein, and
  Seltzer}{Han et~al\mbox{.}}{2017}]%
        {han2017frappuccino}
\bibfield{author}{\bibinfo{person}{Xueyuan Han}, \bibinfo{person}{Thomas
  Pasquier}, \bibinfo{person}{Tanvi Ranjan}, \bibinfo{person}{Mark Goldstein},
  {and} \bibinfo{person}{Margo Seltzer}.} \bibinfo{year}{2017}\natexlab{}.
\newblock \showarticletitle{{Frappuccino: Fault-detection through runtime
  analysis of provenance}}. In \bibinfo{booktitle}{\emph{Workshop on Hot Topics
  in Cloud Computing (HotCloud'17)}}. USENIX.
\newblock


\bibitem[\protect\citeauthoryear{Han, Yu, Pasquier, Li, Rhee, Mickens, Seltzer,
  and Chen}{Han et~al\mbox{.}}{2021}]%
        {han2020sigl}
\bibfield{author}{\bibinfo{person}{Xueyuan Han}, \bibinfo{person}{Xiao Yu},
  \bibinfo{person}{Thomas Pasquier}, \bibinfo{person}{Ding Li},
  \bibinfo{person}{Junghwan Rhee}, \bibinfo{person}{James Mickens},
  \bibinfo{person}{Margo Seltzer}, {and} \bibinfo{person}{Haifeng Chen}.}
  \bibinfo{year}{2021}\natexlab{}.
\newblock \showarticletitle{{SIGL: Securing Software Installations Through Deep
  Graph Learning}}. In \bibinfo{booktitle}{\emph{Security Symposium}}.
  \bibinfo{publisher}{USENIX}.
\newblock


\bibitem[\protect\citeauthoryear{Hassan, Aguse, Aguse, Bates, and Moyer}{Hassan
  et~al\mbox{.}}{2018}]%
        {hassan2018towards}
\bibfield{author}{\bibinfo{person}{Wajih~Ul Hassan}, \bibinfo{person}{Lemay
  Aguse}, \bibinfo{person}{Nuraini Aguse}, \bibinfo{person}{Adam Bates}, {and}
  \bibinfo{person}{Thomas Moyer}.} \bibinfo{year}{2018}\natexlab{}.
\newblock \showarticletitle{{Towards scalable cluster auditing through
  grammatical inference over provenance graphs}}. In
  \bibinfo{booktitle}{\emph{Network and Distributed Systems Security Symposium
  (NDSS'18)}}.
\newblock


\bibitem[\protect\citeauthoryear{Hassan, Bates, and Marino}{Hassan
  et~al\mbox{.}}{2020a}]%
        {hassan2020tactical}
\bibfield{author}{\bibinfo{person}{Wajih~Ul Hassan}, \bibinfo{person}{Adam
  Bates}, {and} \bibinfo{person}{Daniel Marino}.}
  \bibinfo{year}{2020}\natexlab{a}.
\newblock \showarticletitle{{Tactical Provenance Analysis for Endpoint
  Detection and Response Systems}}. In \bibinfo{booktitle}{\emph{Symposium on
  Security and Privacy (S\&P'20)}}. IEEE.
\newblock


\bibitem[\protect\citeauthoryear{Hassan, Noureddine, Datta, and Bates}{Hassan
  et~al\mbox{.}}{2020b}]%
        {hassan2020omegalog}
\bibfield{author}{\bibinfo{person}{Wajih~Ul Hassan},
  \bibinfo{person}{Mohammad~Ali Noureddine}, \bibinfo{person}{Pubali Datta},
  {and} \bibinfo{person}{Adam Bates}.} \bibinfo{year}{2020}\natexlab{b}.
\newblock \showarticletitle{{OmegaLog: High-fidelity attack investigation via
  transparent multi-layer log analysis}}. In \bibinfo{booktitle}{\emph{Network
  and Distributed System Security Symposium}}. Internet Society.
\newblock


\bibitem[\protect\citeauthoryear{Heo}{Heo}{[n.d.]}]%
        {cgroupv2}
\bibfield{author}{\bibinfo{person}{Tejun Heo}.}
  \bibinfo{year}{[n.d.]}\natexlab{}.
\newblock \bibinfo{title}{{Control Group v2}}.
\newblock \bibinfo{howpublished}{online \reviewchange{(accessed \today)}}.
\newblock
\newblock
\shownote{\url{https://www.kernel.org/doc/html/latest/admin-guide/cgroup-v2.html}.}


\bibitem[\protect\citeauthoryear{Jaeger, Edwards, and Zhang}{Jaeger
  et~al\mbox{.}}{2004}]%
        {jaeger2004consistency}
\bibfield{author}{\bibinfo{person}{Trent Jaeger}, \bibinfo{person}{Antony
  Edwards}, {and} \bibinfo{person}{Xiaolan Zhang}.}
  \bibinfo{year}{2004}\natexlab{}.
\newblock \showarticletitle{{Consistency analysis of authorization hook
  placement in the Linux security modules framework}}.
\newblock \bibinfo{journal}{\emph{ACM Transactions on Information and System
  Security (TISSEC)}} \bibinfo{volume}{7}, \bibinfo{number}{2}
  (\bibinfo{year}{2004}), \bibinfo{pages}{175--205}.
\newblock


\bibitem[\protect\citeauthoryear{Jin, Li, Zou, and Yuan}{Jin
  et~al\mbox{.}}{2019}]%
        {jin2019dseom}
\bibfield{author}{\bibinfo{person}{Hai Jin}, \bibinfo{person}{Zhi Li},
  \bibinfo{person}{Deqing Zou}, {and} \bibinfo{person}{Bin Yuan}.}
  \bibinfo{year}{2019}\natexlab{}.
\newblock \showarticletitle{{Dseom: A framework for dynamic security evaluation
  and optimization of MTD in container-based cloud}}.
\newblock \bibinfo{journal}{\emph{IEEE Transactions on Dependable and Secure
  Computing}} (\bibinfo{year}{2019}).
\newblock


\bibitem[\protect\citeauthoryear{Johansen and Schaufler}{Johansen and
  Schaufler}{2017}]%
        {namespacing2017}
\bibfield{author}{\bibinfo{person}{John Johansen} {and} \bibinfo{person}{Casey
  Schaufler}.} \bibinfo{year}{2017}\natexlab{}.
\newblock \showarticletitle{{Namespacing and Stacking the LSM}}. In
  \bibinfo{booktitle}{\emph{Linux Plumbers Conference}}.
\newblock


\bibitem[\protect\citeauthoryear{Katcher}{Katcher}{1997}]%
        {katcher1997postmark}
\bibfield{author}{\bibinfo{person}{Jeffrey Katcher}.}
  \bibinfo{year}{1997}\natexlab{}.
\newblock \bibinfo{booktitle}{\emph{{Postmark: A new file system benchmark}}}.
\newblock \bibinfo{type}{{T}echnical {R}eport}. \bibinfo{institution}{Technical
  Report TR3022, Network Appliance}.
\newblock


\bibitem[\protect\citeauthoryear{Lee, Zhang, and Xu}{Lee et~al\mbox{.}}{2013}]%
        {lee2013high}
\bibfield{author}{\bibinfo{person}{Kyu~Hyung Lee}, \bibinfo{person}{Xiangyu
  Zhang}, {and} \bibinfo{person}{Dongyan Xu}.} \bibinfo{year}{2013}\natexlab{}.
\newblock \showarticletitle{{High Accuracy Attack Provenance via Binary-based
  Execution Partition}}. In \bibinfo{booktitle}{\emph{Network and Distributed
  System Security Symposium (NDSS'13)}}. Internet Society.
\newblock


\bibitem[\protect\citeauthoryear{Liu, Wen, Zhang, Jiang, Xing, and Meng}{Liu
  et~al\mbox{.}}{2019}]%
        {liu2019log2vec}
\bibfield{author}{\bibinfo{person}{Fucheng Liu}, \bibinfo{person}{Yu Wen},
  \bibinfo{person}{Dongxue Zhang}, \bibinfo{person}{Xihe Jiang},
  \bibinfo{person}{Xinyu Xing}, {and} \bibinfo{person}{Dan Meng}.}
  \bibinfo{year}{2019}\natexlab{}.
\newblock \showarticletitle{{Log2vec: a heterogeneous graph embedding based
  approach for detecting cyber threats within enterprise\balance}}. In
  \bibinfo{booktitle}{\emph{Conference on Computer and Communications Security
  (CCS'19)}}. ACM, \bibinfo{pages}{1777--1794}.
\newblock


\bibitem[\protect\citeauthoryear{Manzoor, Milajerdi, and Akoglu}{Manzoor
  et~al\mbox{.}}{2016}]%
        {manzoor2016fast}
\bibfield{author}{\bibinfo{person}{Emaad Manzoor}, \bibinfo{person}{Sadegh~M
  Milajerdi}, {and} \bibinfo{person}{Leman Akoglu}.}
  \bibinfo{year}{2016}\natexlab{}.
\newblock \showarticletitle{{Fast memory-efficient anomaly detection in
  streaming heterogeneous graphs}}. In \bibinfo{booktitle}{\emph{International
  Conference on Knowledge Discovery and Data Mining (KDD'16)}}. ACM,
  \bibinfo{pages}{1035--1044}.
\newblock


\bibitem[\protect\citeauthoryear{McVoy, Staelin, et~al\mbox{.}}{McVoy
  et~al\mbox{.}}{1996}]%
        {mcvoy1996lmbench}
\bibfield{author}{\bibinfo{person}{Larry~W McVoy}, \bibinfo{person}{Carl
  Staelin}, {et~al\mbox{.}}} \bibinfo{year}{1996}\natexlab{}.
\newblock \showarticletitle{{lmbench: Portable Tools for Performance
  Analysis}}. In \bibinfo{booktitle}{\emph{Annual Technical Conference
  (ATC'96)}}. USENIX, \bibinfo{pages}{279--294}.
\newblock


\bibitem[\protect\citeauthoryear{Milajerdi, Eshete, Gjomemo, and
  Venkatakrishnan}{Milajerdi et~al\mbox{.}}{2019a}]%
        {milajerdi2019poirot}
\bibfield{author}{\bibinfo{person}{Sadegh~M. Milajerdi},
  \bibinfo{person}{Birhanu Eshete}, \bibinfo{person}{Rigel Gjomemo}, {and}
  \bibinfo{person}{V.~N. Venkatakrishnan}.} \bibinfo{year}{2019}\natexlab{a}.
\newblock \showarticletitle{{Poirot: Aligning Attack Behavior with Kernel Audit
  Records for Cyber Threat Hunting}}. In \bibinfo{booktitle}{\emph{Conference
  on Computer and Communications Security (CCS'19)}}. \bibinfo{publisher}{ACM}.
\newblock


\bibitem[\protect\citeauthoryear{Milajerdi, Gjomemo, Eshete, Sekar, and
  Venkatakrishnan}{Milajerdi et~al\mbox{.}}{2019b}]%
        {milajerdi2019holmes}
\bibfield{author}{\bibinfo{person}{Sadegh~M Milajerdi}, \bibinfo{person}{Rigel
  Gjomemo}, \bibinfo{person}{Birhanu Eshete}, \bibinfo{person}{Ramachandran
  Sekar}, {and} \bibinfo{person}{VN Venkatakrishnan}.}
  \bibinfo{year}{2019}\natexlab{b}.
\newblock \showarticletitle{{Holmes: real-time apt detection through
  correlation of suspicious information flows}}. In
  \bibinfo{booktitle}{\emph{Symposium on Security and Privacy (S\&P'19)}}.
  IEEE, \bibinfo{pages}{1137--1152}.
\newblock


\bibitem[\protect\citeauthoryear{Morris, Smalley, and Kroah-Hartman}{Morris
  et~al\mbox{.}}{2002}]%
        {morris2002linux}
\bibfield{author}{\bibinfo{person}{James Morris}, \bibinfo{person}{Stephen
  Smalley}, {and} \bibinfo{person}{Greg Kroah-Hartman}.}
  \bibinfo{year}{2002}\natexlab{}.
\newblock \showarticletitle{{Linux Security Modules: General Security Support
  for the Linux Kernel}}. In \bibinfo{booktitle}{\emph{Security Symposium}}.
  \bibinfo{publisher}{USENIX}.
\newblock


\bibitem[\protect\citeauthoryear{Moyer and Gadepally}{Moyer and
  Gadepally}{2016}]%
        {moyer2016high}
\bibfield{author}{\bibinfo{person}{Thomas Moyer} {and} \bibinfo{person}{Vijay
  Gadepally}.} \bibinfo{year}{2016}\natexlab{}.
\newblock \showarticletitle{High-throughput ingest of data provenance records
  into Accumulo}. In \bibinfo{booktitle}{\emph{High Performance Extreme
  Computing Conference (HPEC'16)}}. IEEE, \bibinfo{pages}{1--6}.
\newblock


\bibitem[\protect\citeauthoryear{Muniswamy-Reddy, Holland, Braun, and
  Seltzer}{Muniswamy-Reddy et~al\mbox{.}}{2006}]%
        {muniswamy2006provenance}
\bibfield{author}{\bibinfo{person}{Kiran-Kumar Muniswamy-Reddy},
  \bibinfo{person}{David~A Holland}, \bibinfo{person}{Uri Braun}, {and}
  \bibinfo{person}{Margo Seltzer}.} \bibinfo{year}{2006}\natexlab{}.
\newblock \showarticletitle{{Provenance-aware Storage Systems}}. In
  \bibinfo{booktitle}{\emph{Annual Technical Conference (ATC'06)}}. USENIX,
  \bibinfo{pages}{43--56}.
\newblock


\bibitem[\protect\citeauthoryear{Nakryiko}{Nakryiko}{[n.d.]}]%
        {CORE-2020}
\bibfield{author}{\bibinfo{person}{Andrii Nakryiko}.}
  \bibinfo{year}{[n.d.]}\natexlab{}.
\newblock \bibinfo{title}{{BPF Portability and CO-RE}}.
\newblock \bibinfo{howpublished}{online \reviewchange{(accessed \today)}}.
\newblock
\newblock
\shownote{\url{https://facebookmicrosites.github.io/bpf/blog/2020/02/19/bpf-portability-and-co-re.html}.}


\bibitem[\protect\citeauthoryear{Nelson, Van~Geffen, Torlak, and Wang}{Nelson
  et~al\mbox{.}}{2020}]%
        {nelson2020specification}
\bibfield{author}{\bibinfo{person}{Luke Nelson}, \bibinfo{person}{Jacob
  Van~Geffen}, \bibinfo{person}{Emina Torlak}, {and} \bibinfo{person}{Xi
  Wang}.} \bibinfo{year}{2020}\natexlab{}.
\newblock \showarticletitle{{Specification and verification in the field:
  Applying formal methods to BPF just-in-time compilers in the Linux kernel}}.
  In \bibinfo{booktitle}{\emph{Symposium on Operating Systems Design and
  Implementation (OSDI'20)}}. USENIX, \bibinfo{pages}{41--61}.
\newblock


\bibitem[\protect\citeauthoryear{Pasquier, Han, Goldstein, Moyer, Eyers,
  Seltzer, and Bacon}{Pasquier et~al\mbox{.}}{2017}]%
        {pasquier2017practical-socc2017}
\bibfield{author}{\bibinfo{person}{Thomas Pasquier}, \bibinfo{person}{Xueyuan
  Han}, \bibinfo{person}{Mark Goldstein}, \bibinfo{person}{Thomas Moyer},
  \bibinfo{person}{David Eyers}, \bibinfo{person}{Margo Seltzer}, {and}
  \bibinfo{person}{Jean Bacon}.} \bibinfo{year}{2017}\natexlab{}.
\newblock \showarticletitle{{Practical Whole-System Provenance Capture}}. In
  \bibinfo{booktitle}{\emph{Symposium on Cloud Computing (SoCC'17)}}.
  \bibinfo{publisher}{ACM}.
\newblock


\bibitem[\protect\citeauthoryear{Pasquier, Han, Moyer, Bates, Hermant, Eyers,
  Bacon, and Seltzer}{Pasquier et~al\mbox{.}}{2018}]%
        {pasquier2018ccs}
\bibfield{author}{\bibinfo{person}{Thomas Pasquier}, \bibinfo{person}{Xueyuan
  Han}, \bibinfo{person}{Thomas Moyer}, \bibinfo{person}{Adam Bates},
  \bibinfo{person}{Olivier Hermant}, \bibinfo{person}{David Eyers},
  \bibinfo{person}{Jean Bacon}, {and} \bibinfo{person}{Margo Seltzer}.}
  \bibinfo{year}{2018}\natexlab{}.
\newblock \showarticletitle{{Runtime Analysis of Whole-System Provenance}}. In
  \bibinfo{booktitle}{\emph{Conference on Computer and Communications Security
  (CCS'18)}}. ACM.
\newblock


\bibitem[\protect\citeauthoryear{Pohly, McLaughlin, McDaniel, and Butler}{Pohly
  et~al\mbox{.}}{[n.d.]}]%
        {hifisrc}
\bibfield{author}{\bibinfo{person}{Devin~J Pohly}, \bibinfo{person}{Stephen
  McLaughlin}, \bibinfo{person}{Patrick McDaniel}, {and} \bibinfo{person}{Kevin
  Butler}.} \bibinfo{year}{[n.d.]}\natexlab{}.
\newblock \bibinfo{title}{{Hi-Fi source code}}.
\newblock \bibinfo{howpublished}{online \reviewchange{(accessed \today)}}.
\newblock
\newblock
\shownote{\url{https://github.com/djpohly/linux}.}


\bibitem[\protect\citeauthoryear{Pohly, McLaughlin, McDaniel, and Butler}{Pohly
  et~al\mbox{.}}{2012}]%
        {pohly2012hi}
\bibfield{author}{\bibinfo{person}{Devin~J Pohly}, \bibinfo{person}{Stephen
  McLaughlin}, \bibinfo{person}{Patrick McDaniel}, {and} \bibinfo{person}{Kevin
  Butler}.} \bibinfo{year}{2012}\natexlab{}.
\newblock \showarticletitle{{Hi-Fi: Collecting High-fidelity Whole-system
  Provenance}}. In \bibinfo{booktitle}{\emph{Annual Computer Security
  Applications Conference (ACSAC'12)}}. ACM, \bibinfo{pages}{259--268}.
\newblock


\bibitem[\protect\citeauthoryear{Sailer, Zhang, Jaeger, and Van~Doorn}{Sailer
  et~al\mbox{.}}{2004}]%
        {sailer2004design}
\bibfield{author}{\bibinfo{person}{Reiner Sailer}, \bibinfo{person}{Xiaolan
  Zhang}, \bibinfo{person}{Trent Jaeger}, {and} \bibinfo{person}{Leendert
  Van~Doorn}.} \bibinfo{year}{2004}\natexlab{}.
\newblock \showarticletitle{{Design and Implementation of a TCG-based Integrity
  Measurement Architecture}}. In \bibinfo{booktitle}{\emph{Security
  Symposium}}, Vol.~\bibinfo{volume}{13}. USENIX, \bibinfo{pages}{223--238}.
\newblock


\bibitem[\protect\citeauthoryear{Schreuders, McGill, and Payne}{Schreuders
  et~al\mbox{.}}{2011}]%
        {schreuders2011empowering}
\bibfield{author}{\bibinfo{person}{Z~Cliffe Schreuders}, \bibinfo{person}{Tanya
  McGill}, {and} \bibinfo{person}{Christian Payne}.}
  \bibinfo{year}{2011}\natexlab{}.
\newblock \showarticletitle{{Empowering end users to confine their own
  applications: The results of a usability study comparing SELinux, AppArmor,
  and FBAC-LSM}}.
\newblock \bibinfo{journal}{\emph{ACM Transactions on Information and System
  Security (TISSEC)}} \bibinfo{volume}{14}, \bibinfo{number}{2}
  (\bibinfo{year}{2011}), \bibinfo{pages}{1--28}.
\newblock


\bibitem[\protect\citeauthoryear{Singh}{Singh}{2019}]%
        {krsi_lwn}
\bibfield{author}{\bibinfo{person}{KP Singh}.} \bibinfo{year}{2019}\natexlab{}.
\newblock \bibinfo{title}{{Kernel Runtime Security Instrumentation}}.
\newblock \bibinfo{howpublished}{online \reviewchange{(accessed \today)}}.
\newblock
\newblock
\shownote{\url{https://lwn.net/Articles/798918/}.}


\bibitem[\protect\citeauthoryear{Smalley, Vance, and Salamon}{Smalley
  et~al\mbox{.}}{2001}]%
        {smalley2001implementing}
\bibfield{author}{\bibinfo{person}{Stephen Smalley}, \bibinfo{person}{Chris
  Vance}, {and} \bibinfo{person}{Wayne Salamon}.}
  \bibinfo{year}{2001}\natexlab{}.
\newblock \showarticletitle{{Implementing SELinux as a Linux security module}}.
\newblock \bibinfo{journal}{\emph{NAI Labs Report}} \bibinfo{volume}{1},
  \bibinfo{number}{43} (\bibinfo{year}{2001}), \bibinfo{pages}{139}.
\newblock


\bibitem[\protect\citeauthoryear{Soltesz, P{\"o}tzl, Fiuczynski, Bavier, and
  Peterson}{Soltesz et~al\mbox{.}}{2007}]%
        {soltesz2007container}
\bibfield{author}{\bibinfo{person}{Stephen Soltesz}, \bibinfo{person}{Herbert
  P{\"o}tzl}, \bibinfo{person}{Marc~E Fiuczynski}, \bibinfo{person}{Andy
  Bavier}, {and} \bibinfo{person}{Larry Peterson}.}
  \bibinfo{year}{2007}\natexlab{}.
\newblock \showarticletitle{{Container-based operating system virtualization: a
  scalable, high-performance alternative to hypervisors}}. In
  \bibinfo{booktitle}{\emph{European Conference on Computer Systems
  (EuroSys'07)}}. ACM, \bibinfo{pages}{275--287}.
\newblock


\bibitem[\protect\citeauthoryear{Sun, Safford, Zohar, Pendarakis, Gu, and
  Jaeger}{Sun et~al\mbox{.}}{2018}]%
        {sun2018security}
\bibfield{author}{\bibinfo{person}{Yuqiong Sun}, \bibinfo{person}{David
  Safford}, \bibinfo{person}{Mimi Zohar}, \bibinfo{person}{Dimitrios
  Pendarakis}, \bibinfo{person}{Zhongshu Gu}, {and} \bibinfo{person}{Trent
  Jaeger}.} \bibinfo{year}{2018}\natexlab{}.
\newblock \showarticletitle{{Security namespace: making linux security
  frameworks available to containers}}. In \bibinfo{booktitle}{\emph{Security
  Symposium}}. USENIX.
\newblock


\bibitem[\protect\citeauthoryear{Tang, Li, Li, Zhang, Jee, Xiao, Wu, Rhee, Xu,
  and Li}{Tang et~al\mbox{.}}{2018}]%
        {Tang:2018:NTB:3243734.3243763}
\bibfield{author}{\bibinfo{person}{Yutao Tang}, \bibinfo{person}{Ding Li},
  \bibinfo{person}{Zhichun Li}, \bibinfo{person}{Mu Zhang},
  \bibinfo{person}{Kangkook Jee}, \bibinfo{person}{Xusheng Xiao},
  \bibinfo{person}{Zhenyu Wu}, \bibinfo{person}{Junghwan Rhee},
  \bibinfo{person}{Fengyuan Xu}, {and} \bibinfo{person}{Qun Li}.}
  \bibinfo{year}{2018}\natexlab{}.
\newblock \showarticletitle{{NodeMerge: Template Based Efficient Data Reduction
  For Big-Data Causality Analysis}}. In \bibinfo{booktitle}{\emph{Conference on
  Computer and Communications Security (CCS'18)}}. ACM,
  \bibinfo{pages}{1324--1337}.
\newblock


\bibitem[\protect\citeauthoryear{Tian, Hernandez, Choi, Frost, Johnson, and
  Butler}{Tian et~al\mbox{.}}{2019}]%
        {tian2019lbm}
\bibfield{author}{\bibinfo{person}{Dave~Jing Tian}, \bibinfo{person}{Grant
  Hernandez}, \bibinfo{person}{Joseph~I Choi}, \bibinfo{person}{Vanessa Frost},
  \bibinfo{person}{Peter~C Johnson}, {and} \bibinfo{person}{Kevin~RB Butler}.}
  \bibinfo{year}{2019}\natexlab{}.
\newblock \showarticletitle{{LBM: a security framework for peripherals within
  the linux kernel}}. In \bibinfo{booktitle}{\emph{Symposium on Security and
  Privacy (S\&P'19)}}. IEEE, \bibinfo{pages}{967--984}.
\newblock


\bibitem[\protect\citeauthoryear{Torkura, Sukmana, and Meinel}{Torkura
  et~al\mbox{.}}{2017}]%
        {torkura2017integrating}
\bibfield{author}{\bibinfo{person}{Kennedy~A Torkura},
  \bibinfo{person}{Muhammad~IH Sukmana}, {and} \bibinfo{person}{Christoph
  Meinel}.} \bibinfo{year}{2017}\natexlab{}.
\newblock \showarticletitle{{Integrating continuous security assessments in
  microservices and cloud native applications}}. In
  \bibinfo{booktitle}{\emph{International Conference on Utility and Cloud
  Computing (UCC'17)}}. IEEE/ACM, \bibinfo{pages}{171--180}.
\newblock


\bibitem[\protect\citeauthoryear{Veritis}{Veritis}{2019}]%
        {containerstate}
\bibfield{author}{\bibinfo{person}{Veritis}.} \bibinfo{year}{2019}\natexlab{}.
\newblock \bibinfo{title}{{State of Containers Report 2019: ‘Security’
  Remains A Challenge!}}
\newblock \bibinfo{howpublished}{online \reviewchange{(accessed \today)}}.
\newblock
\newblock
\shownote{\url{https://www.veritis.com/blog/state-of-containers-report-2019-security-remains-a-challenge/}.}


\bibitem[\protect\citeauthoryear{Wang, Hassan, Li, Jee, Yu, Zou, Rhee, Chen,
  Cheng, Gunter, and Chen}{Wang et~al\mbox{.}}{2020}]%
        {wang2020you}
\bibfield{author}{\bibinfo{person}{Qi Wang}, \bibinfo{person}{Wajih~Ul Hassan},
  \bibinfo{person}{Ding Li}, \bibinfo{person}{Kangkook Jee},
  \bibinfo{person}{Xiao Yu}, \bibinfo{person}{Kexuan Zou},
  \bibinfo{person}{Junghwan Rhee}, \bibinfo{person}{Zhengzhang Chen},
  \bibinfo{person}{Wei Cheng}, \bibinfo{person}{Carl~A. Gunter}, {and}
  \bibinfo{person}{Haifeng Chen}.} \bibinfo{year}{2020}\natexlab{}.
\newblock \showarticletitle{{You Are What You Do: Hunting Stealthy Malware via
  Data Provenance Analysis}}. In \bibinfo{booktitle}{\emph{Network and
  Distributed System Security (NDSS'20)}}. Internet Society.
\newblock


\bibitem[\protect\citeauthoryear{Watson}{Watson}{2007}]%
        {watson2007exploiting}
\bibfield{author}{\bibinfo{person}{Robert~NM Watson}.}
  \bibinfo{year}{2007}\natexlab{}.
\newblock \showarticletitle{{Exploiting Concurrency Vulnerabilities in System
  Call Wrappers}}.
\newblock \bibinfo{journal}{\emph{Workshop on Offensive Technologies
  (WOOT'07)}}  \bibinfo{volume}{7} (\bibinfo{year}{2007}),
  \bibinfo{pages}{1--8}.
\newblock


\bibitem[\protect\citeauthoryear{Watson}{Watson}{2013}]%
        {watson2013decade}
\bibfield{author}{\bibinfo{person}{Robert~NM Watson}.}
  \bibinfo{year}{2013}\natexlab{}.
\newblock \showarticletitle{{A decade of OS access-control extensibility}}.
\newblock \bibinfo{journal}{\emph{ACM Queue}} \bibinfo{volume}{11},
  \bibinfo{number}{1} (\bibinfo{year}{2013}), \bibinfo{pages}{20--41}.
\newblock


\bibitem[\protect\citeauthoryear{Zhang, Liu, and Jaeger}{Zhang
  et~al\mbox{.}}{2021}]%
        {zhang2021analyzing}
\bibfield{author}{\bibinfo{person}{Wenhui Zhang}, \bibinfo{person}{Peng Liu},
  {and} \bibinfo{person}{Trent Jaeger}.} \bibinfo{year}{2021}\natexlab{}.
\newblock \showarticletitle{{Analyzing the Overhead of File Protection by Linux
  Security Modules}}. In \bibinfo{booktitle}{\emph{Asia Conference on Computer
  and Communications Security (AsiaCCS'21)}}. ACM, \bibinfo{pages}{393--406}.
\newblock


\end{thebibliography}
\end{document}